\documentclass[11pt,a4paper]{article}
\usepackage{jheppub}
\usepackage[T1]{fontenc}
\usepackage{amsmath,amssymb}
\usepackage{xcolor}
\usepackage{booktabs}
\usepackage{placeins}
\usepackage[normalem]{ulem}

\newcommand{\ga}{g_{a\gamma}}
\newcommand{\gH}{g_{H\gamma}}
\newcommand{\gL}{g_{L\gamma}}

\title{Dark Matter Assisted Axion-Magnetic Resonance}
\author[a,b,1]{Chen Sun,}
\author[a,c,d,1]{Si-Yu Wu}
\author[e,f,1]{and Seokhoon Yun\note{Corresponding authors.}}
\affiliation[a]{School of Fundamental Physics and Mathematical Sciences,\\
Hangzhou Institute for Advanced Study, University of Chinese Academy of Sciences,\\
Hangzhou 310024, China}
\affiliation[b]{International Centre for Theoretical Physics Asia-Pacific, Beijing / Hangzhou, China}
\affiliation[c]{University of Chinese Academy of Sciences, Beijing 100049, China}
\affiliation[d]{Institute of Theoretical Physics, Chinese Academy of Sciences,\\
Beijing 100190, China}
\affiliation[e]{Department of Physics, Kyungpook National University,\\
Daegu 41566, Korea}
\affiliation[f]{Center for Theoretical Physics of the Universe, Institute for Basic Science (IBS), Daejeon, 34126, Korea}
\emailAdd{chen.sun@ucas.ac.cn}
\emailAdd{wusiyu25@mails.ucas.ac.cn}
\emailAdd{seokhoon.yun@knu.ac.kr}

\abstract{An oscillating axion dark matter (DM) background can enhance the
conversion of another relativistic axion into photons when traversing a
magnetic field. The DM background rotates photon polarization relative the magnetic orientation, which compensates the photon's and the propagating axion's dispersion relation mismatch.
We relate this mechanism to Axion Magnetic Resonance (AMR) and apply it to
stellar axions in the starburst galaxy M82. We analyze its enhancement to the resulting X-ray signals while accounting for the finite magnetic domain size, dark matter coherence length, DM density morphology, plasma profiles,
and the detector response. We derive strong constraints using existing X-ray data from NuSTAR. We show that this resonant conversion can lead to $\mathcal O(10^2)$ enhancement in the photon-axion conversion probability, improving the upper limits on axion's
photon coupling over part of the mass range where ordinary conversion is
suppressed. We also show that this type of resonance predicts distinctive features in the hard X-ray spectrum, which is moderately favored by the data over the background only model, yet it is currently still consistent with statistical fluctuations.}

\begin{document}
\maketitle
\flushbottom

\section{Introduction}
\label{sec:introduction}

Axions are light pseudoscalar particles associated with approximate shift
symmetries. The QCD axion arises from the Peccei--Quinn solution to the
strong CP problem~\cite{Peccei:1977StrongCP,Weinberg:1978Axion,Wilczek:1978Axion}.
More general axion-like particles (ALPs) need not obey the QCD relation
between their masses and couplings. Such fields arise naturally in string
compactifications, which can contain many axions with masses spanning a
wide range~\cite{Svrcek:2006StringAxions,Arvanitaki:2009Axiverse}. This
axiverse motivates studying the observable consequences of several axion
species coexisting in the same environment.

Axions are also compelling dark matter (DM) candidates. Their weak
couplings can make them cosmologically long-lived, while vacuum
misalignment produces coherent oscillations that behave as cold matter,
even for very small particle masses~\cite{Preskill:1983AxionDM,Abbott:1983AxionDM,Dine:1983AxionDM,Hui:2017ULDM}.
For the QCD axion, this connects DM to an independently motivated solution
of the strong CP problem. The presence of many axions does not require
comparable relic abundances: a suitable inflationary history can suppress
most species while allowing one to dominate the axion contribution to
DM~\cite{Graham:2018Stochastic,Reig:2021Axiverse,Murai:2023Axiverse}.

In this work, we extend the study of axion magnetic resonance
(AMR)~\cite{Seong:2023AMR,Seong:2024Helioscopes} to axion conversion in
galactic DM backgrounds. The oscillating background provides a periodic
drive that can resonantly enhance the conversion of another, relativistic
axion species into photons in an external magnetic field. Within a
coherent region, the conversion probability can increase by orders of
magnitude, making the coexisting axion species easier to detect.

The coherent oscillations of axion DM have been studied in several different and related contexts. An ultralight axion DM component has a large occupation
number and can be
described locally as a coherent classical field oscillating at a frequency
set by its mass. Its oscillating pressure produces a time-dependent
gravitational potential that can be probed through pulsar
timing~\cite{Khmelnitsky:2013Pulsar}, and the gravitational perturbation
can resonantly affect binary orbits~\cite{Blas:2016Binary}. Through its
coupling to photons, the axion background also rotates the polarization
of propagating light~\cite{Liu:2019Pulsar,PPTA:2024Polarization}.
Recent work has shown that this periodic birefringence can drive resonant
photon--axion conversion in a magnetic
field~\cite{Yao:2026DM,Yao:2026Sidebands}.

The connection to AMR provides a useful physical
picture. 
{In a uniform magnetic field, the axion–photon momentum mismatch causes conversion amplitudes generated at different positions to partially cancel once the accumulated relative phase becomes appreciable across the conversion region.
A spatially rotating magnetic field can compensate this phase mismatch and allow the conversion amplitudes to accumulate constructively~\cite{Seong:2023AMR,Seong:2024Helioscopes}.}
An oscillating axion background supplies a related modulation by rotating
the photon polarization relative to the magnetic field. In a photon basis
that follows this rotation, the effect appears as a varying magnetic-field
direction. 
Within {a region over which the magnetic-field direction and the DM oscillation phase remain correlated}, conversion is resonantly enhanced
when a harmonic of the background oscillation compensates the
axion--photon momentum mismatch. The DM axion mass sets the modulation
frequency and, together with the propagating axion mass, selects the
resonant photon energies. The modulation amplitude depends on the DM
density and the light axion's mass and photon coupling.

For a concrete application, we consider two distinct axion species. An ultralight field $a_L$ supplies a fraction of the DM, while a
heavier field $a_H$ has a negligible cosmological relic abundance but is
produced in stars. The stellar axions convert into X rays as they
propagate through the galactic magnetic field. We apply this scenario to
the nearby starburst galaxy M82, whose stellar population and NuSTAR
observations already provide a sensitive test of ordinary axion
conversion~\cite{Ning:2024M82}. Using realistic modeling of the starburst galaxy, we calculate how the resonance is affected
by spatially varying galactic profiles, finite magnetic and DM coherence,
and energy averaging, and compare the predicted counts with the NuSTAR
spectrum. Under the adopted source and halo assumptions, the DM background
strengthens the M82 limits in a mass range where ordinary conversion is
strongly suppressed by the axion mass. We also search for the associated spectral
features and assess their statistical significance. This provides an
observational test of the combined stellar-axion and axion-DM system.

Section~\ref{sec:mechanism} develops the resonance mechanism and its
connection to AMR. Section~\ref{sec:m82} describes the M82 model and signal
calculation. Sections~\ref{sec:results}
 present the constraints and spectral-feature
search, and we conclude in Section~\ref{sec:conclusions}.

\section{Axion magnetic resonance}
\label{sec:mechanism}

\subsection{Phase matching lost and regained}
\label{sec:coherence-limit}
\paragraph{Generic formalism.}
Axion--photon conversion in a macroscopic magnetic field relies on the
coherent addition of amplitudes generated along the propagation path.
For a propagating axion $a$ with photon coupling $\ga$, the interaction
\begin{equation}
  \mathcal L_{\rm int}
  =-\frac{\ga}{4}\,aF_{\mu\nu}\widetilde F^{\mu\nu}
  =\ga\,a\,\mathbf E\cdot\mathbf B
  \label{eq:axion-photon-interaction}
\end{equation}
couples an axion to a photon whose polarization vector is along
the external magnetic field~\cite{Sikivie:1983ip,Raffelt:1987im}.
In what follows, we consider an axion field
of mass $m_a$ and energy $\omega$ propagating in the $z$ direction. We parameterize the
photon dispersion by an effective mass $m_\gamma$, which can be provided by, say, plasma frequency of a transverse photon mode. Its momentum is fixed by the usual dispersion relation
$k_\gamma^2=\omega^2-m_\gamma^2$. We assume $\omega\gg m_a,m_\gamma$.

For two degenerate transverse photon polarizations and a prescribed
static transverse field $(B_x,B_y)$, the slowly varying amplitudes obey
\begin{equation}
  i\frac{d}{dz}
  \begin{pmatrix} A_x\\ A_y\\ a\end{pmatrix}
  =
  \begin{pmatrix}
    \Delta_{a\gamma}/2 & 0 & \ga B_x/2\\
    0 & \Delta_{a\gamma}/2 & \ga B_y/2\\
    \ga B_x/2 & \ga B_y/2 & -\Delta_{a\gamma}/2
  \end{pmatrix}
  \begin{pmatrix} A_x\\ A_y\\ a\end{pmatrix},
  \label{eq:ordinary-propagation}
\end{equation}
where a common propagation phase has been removed. The magnetic mixing
strength is $\Delta_B\equiv\ga B_T/2$, where
$B_T\equiv\sqrt{B_x^2+B_y^2}$ is the transverse field magnitude.
We assume that the
background varies slowly on the scale $\omega^{-1}$.
Expanding the wave numbers at fixed energy gives
$k_a\simeq \omega-m_a^2/(2\omega)$ and
$k_\gamma\simeq \omega-m_\gamma^2/(2\omega)$, and hence the mismatch
\begin{equation}
  \Delta_{a\gamma}=k_\gamma-k_a
  \simeq\frac{m_a^2-m_\gamma^2}{2\omega}.
  \label{eq:momentum-mismatch}
\end{equation}
Thus a phase difference
accumulates even in a homogeneous medium, without any spread of particle
velocities.

For an initially pure axion, integrating
Eq.~\eqref{eq:ordinary-propagation} to leading order in $\ga$ gives
\begin{equation}
  P_{a\to\gamma}(L)
  \simeq \frac{\ga^2}{4}
  \sum_{\lambda=x,y}
  \left|
    \int_0^L dz\,B_\lambda(z)
    \exp\!\left[i\int_0^z dz'\,\Delta_{a\gamma}(z')\right]
  \right|^2 .
  \label{eq:coherent-conversion-integral}
\end{equation}
The integral adds amplitudes before taking their absolute square.  For constant $\Delta_{a\gamma}$ and a uniform transverse field of magnitude
$B_T$, it reduces to
\begin{equation}
  P_{a\to\gamma}(L)
  \simeq
  (\Delta_B L)^2
  \operatorname{sinc}^2\!\left(\frac{\Delta_{a\gamma}L}{2}\right),
  \qquad
  \operatorname{sinc}x\equiv\frac{\sin x}{x}.
  \label{eq:uniform-coherence}
\end{equation}
When $|\Delta_{a\gamma}|L\ll1$, 
{the conversion amplitudes generated along the path add approximately in phase,} and the probability
grows as $L^2$ within the perturbative regime.
Once $|\Delta_{a\gamma}|L$ becomes large, different portions of the path contribute with opposing phases, and the probability oscillates instead of continuing to grow as $L^2$.

\paragraph{Examples of phase matching.}
The conversion integral also suggests ways to recover the constructive buildup:
one should reduce the propagation mismatch $\Delta_{a\gamma}$ by a compensating mechanism. We list three examples to illustrate such possibilities.
\begin{itemize}
  \item \textit{Plasma frequency.}
  In an isotropic plasma, the effective photon mass is
  $m_\gamma=\omega_{\rm pl}$. Tuning $\omega_{\rm pl}\simeq m_a$ makes
  $\Delta_{a\gamma}$ small, 
  {allowing conversion amplitudes to accumulate constructively over the conversion region.}
  This is the principle behind buffer-gas axion
  helioscopes~\cite{vanBibber:1988ge}.

  \item \textit{Cavity and waveguide modes.}
  Transverse confinement provides a geometric contribution to the photon
  dispersion~\cite{Batllori:2024WISPFI}. For example, in an evacuated
  waveguide with perfectly conducting walls, the transverse wave number
  $k_\perp$ takes discrete values fixed by the cross-section and mode.
  The longitudinal dispersion is $\omega^2=k_\gamma^2+k_\perp^2$, so the effective
  mass for longitudinal propagation is $m_\gamma=k_\perp$. For an axion
  incident along the guide,
  \begin{equation}
    \Delta_{a\gamma}=k_\gamma-k_a
    =\sqrt{\omega^2-k_\perp^2}-\sqrt{\omega^2-m_a^2}
    \simeq\frac{m_a^2-k_\perp^2}{2\omega},
    \label{eq:waveguide-phase-matching}
  \end{equation}
  where the last expression assumes $\omega\gg m_a,k_\perp$.
  Choosing $k_\perp\simeq m_a$ restores longitudinal phase matching.

  \item \textit{Magnetic-field modulation.}
  A rotating magnetic field can supply the missing spatial phase.
  Consider axion propagation in vacuum through a transverse field of
  constant magnitude whose direction rotates at a constant, signed
  spatial rate $\dot\theta\equiv d\theta/dz$, so that
  $\theta(z)=\dot\theta z$. Its circular components contribute
  factors $e^{\mp i\dot\theta z}$ to
  Eq.~\eqref{eq:coherent-conversion-integral}. Since the vacuum mismatch
  is $\Delta_{a\gamma}\simeq m_a^2/(2\omega)$, the two circular photon
  channels have phase mismatches
  \begin{equation}
    \Delta_{a\gamma,\pm}=\Delta_{a\gamma}\mp\dot\theta
    \simeq\frac{m_a^2}{2\omega}\mp\dot\theta .
    \label{eq:modulation-phase-mismatch}
  \end{equation}
  For a fixed axion mass $m_a$ and energy $\omega$, tuning the rotation
  rate to
  \begin{equation}
    \dot\theta=\pm\Delta_{a\gamma}
    \simeq\pm\frac{m_a^2}{2\omega}
    \label{eq:modulation-resonance}
  \end{equation}
  cancels the {phase mismatch in one circular polarization channel} and restores coherent
  buildup~\cite{Seong:2023AMR,Seong:2024Helioscopes}. The sign selects which circular
  polarization is resonant.
\end{itemize}
The following subsections focus on the last mechanism, first for a
review of AMR using the helical magnetic field, then the main focus on an oscillating axion DM background.

\subsection{Brief review on axion-magnetic resonance (AMR)}
\label{sec:helical-field}
A simple realization of \textit{static} magnetic phase matching is a transverse field
of constant magnitude whose direction rotates uniformly along the
propagation path~\cite{Seong:2023AMR}:
\begin{equation}
  \mathbf B_T(z)
  =B_T\bigl(\sin\theta(z),\,\cos\theta(z),\,0\bigr),
  \qquad \theta(z)=\dot\theta z .
  \label{eq:helical-field-profile}
\end{equation}
Here $\dot\theta$ is a constant, signed spatial rotation rate; a full turn
corresponds to a distance $2\pi/|\dot\theta|$.
We take $\Delta_{a\gamma}$ and $B_T$ to be constant to
isolate the effect of the helical magnetic profile.
Substituting this field into Eq.~\eqref{eq:ordinary-propagation} gives
\begin{equation}
  i\frac{d}{dz}
  \begin{pmatrix} A_x\\ A_y\\ a\end{pmatrix}
  =
  \begin{pmatrix}
    \Delta_{a\gamma}/2 & 0 & \ga B_T\sin\theta/2\\
    0 & \Delta_{a\gamma}/2 & \ga B_T\cos\theta/2\\
    \ga B_T\sin\theta/2 & \ga B_T\cos\theta/2 & -\Delta_{a\gamma}/2
  \end{pmatrix}
  \begin{pmatrix} A_x\\ A_y\\ a\end{pmatrix}.
  \label{eq:helical-laboratory-hamiltonian}
\end{equation}

The mechanism is particularly transparent in a photon basis that follows
the magnetic field. Define
\begin{equation}
  \begin{aligned}
    A_\perp&=\cos\theta\,A_x-\sin\theta\,A_y,\\
    A_\parallel&=\sin\theta\,A_x+\cos\theta\,A_y .
  \end{aligned}
  \label{eq:rotating-photon-basis}
\end{equation}
Only $A_\parallel$ couples directly to the axion, but differentiating
the rotating basis generates photon mixing proportional to $\dot\theta$.
In this basis, Eq.~\eqref{eq:helical-laboratory-hamiltonian} becomes
\begin{equation}
  i\frac{d}{dz}
  \begin{pmatrix} A_\perp\\ A_\parallel\\ a\end{pmatrix}
  =
  \begin{pmatrix}
    \Delta_{a\gamma}/2 & -i\dot\theta & 0\\
    i\dot\theta & \Delta_{a\gamma}/2 & \ga B_T/2\\
    0 & \ga B_T/2 & -\Delta_{a\gamma}/2
  \end{pmatrix}
  \begin{pmatrix} A_\perp\\ A_\parallel\\ a\end{pmatrix}.
  \label{eq:helical-rotating-hamiltonian}
\end{equation}
The change of basis moves the rotation from the axion--photon mixing
terms into the photon block. This form will allow a direct comparison
with DM-induced photon polarization rotation later in
Section~\ref{sec:dm-background}.

The circular combinations
$A_\pm=(A_\parallel\mp iA_\perp)/\sqrt{2}$ diagonalize the photon block.
Their diagonal entries are $\Delta_{a\gamma}/2\mp\dot\theta$, and each couples to the axion
with strength $\ga B_T/(2\sqrt{2})$. The relative phase mismatches are
therefore $\Delta_{a\gamma,\pm}=\Delta_{a\gamma}\mp\dot\theta$, as in
Eq.~\eqref{eq:modulation-phase-mismatch}. One branch becomes phase matched when
\begin{equation}
  |\dot\theta|=|\Delta_{a\gamma}|
  \simeq\frac{|m_a^2-m_\gamma^2|}{2\omega}.
  \label{eq:helical-resonance}
\end{equation}
The conversion probability follows directly by substituting
Eq.~\eqref{eq:helical-field-profile} into
Eq.~\eqref{eq:coherent-conversion-integral}:
\begin{equation}
  P_{a\to\gamma}(L)
  \simeq
  \frac{(\ga B_TL)^2}{8}
  \sum_{\pm}
  \operatorname{sinc}^2\!\left(\frac{\Delta_{a\gamma,\pm} L}{2}\right).
  \label{eq:helical-weak-conversion}
\end{equation}
A sufficient
condition for this expression is $|\ga B_TL|\ll1$. It reduces to
Eq.~\eqref{eq:uniform-coherence} when $\dot\theta=0$.
At a resolved resonance, $|\dot\theta|=|\Delta_{a\gamma}|$ and $|\Delta_{a\gamma}|L\gg1$, the resonant
term gives $P_{a\to\gamma}\simeq(\ga B_TL)^2/8$.
Thus the $L^2$ buildup is recovered with half the probability of an
ideally phase-matched fixed field, because only one circular channel
is resonant. A finite path tolerates detuning
$|\Delta_{a\gamma,\pm}|L\lesssim1$; increasing $L$ raises the peak and narrows
the resonance.

The original AMR proposal applied this mechanism to
light-shining-through-walls experiments, including proposed helical-field
configurations for ALPS~II~\cite{Seong:2023AMR}.
For a monochromatic beam, varying the helix pitch scans the resonant
axion mass. A discrete realization uses an array of dipole
magnets with a fixed angular increment between neighboring field
directions about the beam axis. Scanning this
angle also gives a substantial projected sensitivity gain for ALPS~II
using the HERA dipole magnets~\cite{Seong:2023AMR}.
The helioscope follow-up instead exploits the continuous
solar-axion spectrum~\cite{Seong:2024Helioscopes}: a fixed pitch selects
the resonant energy
$\omega\simeq|m_a^2-m_\gamma^2|/(2|\dot\theta|)$, allowing different masses
to be probed at different energies. It studies the resulting sensitivity
for proposed helical-field versions of CAST and IAXO.

There are other studies on AMR-like phenomena by leveraging a non-trivial  magnetic profile. To name a few examples, Van Bibber et al.~\cite{VanBibber:1987rq} proposed an alternating-polarity
magnetic array (a wiggler) to compensate the axion--photon phase mismatch
and extend the accessible mass range, which could be more easily achievable experimentally albeit has a weaker enhancement. Wang and Lai~\cite{Wang:2015dil}
derived the rotating-basis propagation equations for a continuously
varying magnetic-field direction. Their weak-conversion amplitude
contains the same phase-matching condition as
Eq.~\eqref{eq:helical-resonance}, although their analytic discussion
focused on rotation rates much larger than the axion--photon
wave-number mismatch, hence, gave an effective suppression to the conversion than the resonant enhancement.
Spatially harmonic~\cite{Zarei:2019sva} and
time-dependent~\cite{Sharifian:2021vsg} magnetic fields have also been
proposed to enhance ALP-induced optical birefringence in cavities.
The modulation supplies momentum or energy, respectively, increasing
the phase difference between photon polarizations. While they focus on the optical phase shift, we compute the conversion probability.

\subsection{Dark matter background induced resonance}
\label{sec:dm-background}

An axion background that varies along the photon trajectory rotates
linear polarization through birefringence~{\cite{Liu:2019Pulsar,Yao:2026DM}}.
This polarization rotation has been studied as a probe of axion background such as axion DM, axion strings, and axion clouds around a black hole, \textit{e.g.} \cite{Fedderke:2019CMB,Liu:2019Pulsar,Agrawal:2019Strings,McDonald:2019Optics,McDonald:2020Bending,Chen:2019EHT,Chen:2021EHT,Chen:2022Tomography,Jain:2021Strings,PPTA:2024Polarization}.

In the presence of an external magnetic field and a DM background, it generates a relative rotation between the photon's polarization vector and the external magnetic field. While being studied in the literature~\cite{Yao:2026DM,Yao:2026Sidebands}, what has been underappreciated in the literature is the close resemblance of axion-magnetic resonance~\cite{Seong:2023AMR,Seong:2024Helioscopes} and the axion background driven resonance, as well as the actual application to realistic observation scenarios to enhance our sensitivity to probe axions. 

In this work, we show with a concrete example how this resonance can manifest itself in galaxies, such as the starburst galaxy M82, to help us better probe parameter space that we have not been able to reach before. 

To make a concrete example, let us first do an order-of-magnitude estimate. We now introduce a lighter DM species
$a_L$, with mass $m_L$ and photon coupling $\gL$. This will be our oscillating DM background. This leads to birefringence when photons propagate through it. 
There is a second axion species, $a_H$, that mixes with photons in a magnetic background at a coupling strength $\gH$. We will work in the scenario of $m_L\ll m_H$.
%
Consider a constant field $\mathbf B_T=B_T(0,1,0)$.
The interaction $-\gL a_LF_{\mu\nu}\widetilde F^{\mu\nu}/4$
generates the polarization rotation
\begin{equation}
  \dot\theta=\frac{\gL}{2}\frac{da_L}{dz},
  \qquad
  \theta(z)=\frac{\gL}{2}a_L(z),
  \label{eq:dm-rotation-angle}
\end{equation}
up to an irrelevant constant angle. Here $a_L(z)$ denotes the
background sampled along the relativistic ray $t=t_0+z$, so
$da_L/dz=(\partial_t+\partial_z)a_L$ includes its time variation.
Using this rotation rate, the propagation equation becomes
\begin{equation}
  i\frac{d}{dz}
  \begin{pmatrix} A_x\\ A_y\\ a_H\end{pmatrix}
  =
  \begin{pmatrix}
    \Delta_{a_H\gamma}/2 & -i\dot\theta & 0\\
    i\dot\theta & \Delta_{a_H\gamma}/2 & \gH B_T/2\\
    0 & \gH B_T/2 & -\Delta_{a_H\gamma}/2
  \end{pmatrix}
  \begin{pmatrix} A_x\\ A_y\\ a_H\end{pmatrix}.
  \label{eq:dm-background-propagation}
\end{equation}
One can easily notice that it shares the same mathematical form as Eq.~\eqref{eq:helical-rotating-hamiltonian},
with the effective magnetic field helical rate now determined by the light axion background.
We stress that the effective rotation comes from the time-dependent axion DM background as we started only with a static homogeneous magnetic profile. We will postpone the actual morphology of the magnetic field to better isolate the DM induced effects. 

\paragraph{Quality of resonance. }
We now focus on a coherent region of the light-axion DM background,
with density $\rho_{\rm DM}$.
Neglecting the DM velocity at leading order -- which we will fix in next sections -- the background along
the ray is
\begin{equation}
  a_L(z)=\frac{\sqrt{2\rho_{\rm DM}}}{m_L}\cos(m_Lz+\phi),
  \label{eq:dm-background-field}
\end{equation}
where $\phi$ is the initial phase. 
For constant $\Delta_{a_H\gamma}$ and $B_T$,
the photon amplitudes at leading order in $\gH$ contain the phases
$e^{i\Delta_{a_H\gamma}z\pm i\theta(z)}$, where $\Delta_{a_H\gamma}z$ comes from the $a_H$-$\gamma$ Rabi cycle and $\theta(z)$ part comes from the $a_L$ induced birefringence.  
The Jacobi--Anger expansion of the latter reads 
\begin{equation}
  e^{\pm i\theta(z)}
  =\sum_{n=-\infty}^{\infty}(\pm i)^n
  J_n\!\left(\frac{\gL\sqrt{2\rho_{\rm DM}}}{2m_L}\right)
  e^{-in(m_Lz+\phi)}
  \label{eq:dm-jacobi-anger}
\end{equation}
exhibits harmonics spaced by $m_L$, where $J_n$ is a Bessel function
of the first kind. Integrating each harmonic over the path gives a
finite-length sinc factor. Averaging over a uniformly distributed
initial phase $\phi$ removes interference between different harmonics,\footnote{We note that for a fixed $\phi$, the harmonic amplitudes must instead be added
before squaring. In other words, in one coherent DM domain, multiple bands can interfere with each other. }
giving
\begin{equation}
  \left\langle P_{a_H\to\gamma}(L)\right\rangle_\phi
  \simeq
  \left(\frac{\gH B_TL}{2}\right)^2
  \sum_{n=-\infty}^{\infty}
  J_n^2\!\left(\frac{\gL\sqrt{2\rho_{\rm DM}}}{2m_L}\right)
  \operatorname{sinc}^2\!\left[\frac{(\Delta_{a_H\gamma}-nm_L)L}{2}\right].
  \label{eq:dm-harmonic-probability}
\end{equation}
This is leading order in $\gH$, but retains the full rotation angle.
 Resonance occurs at $\Delta_{a_H\gamma}\simeq nm_L$, with
$|\Delta_{a_H\gamma}-nm_L|L\lesssim1$ setting the band locations. The resonance height is determined by 
the rotation amplitude, which in turn fixed by $g_{L\gamma}\sqrt {\rho_{\rm DM}}/m_L$.
For $\Delta_{a_H\gamma}>0$, the first band is centered at $\Delta_{a_H\gamma}=m_L$.   In the case of $m_L L\gg1$, different harmonics are well-separated. The first resonance band has a peak probability given by
\begin{equation}
  P^{\rm res}_{a_H \to \gamma}
  \simeq
  \left(\frac{\gH B_TL}{2}\right)^2
  J_1^2\!\left(\frac{\gL\sqrt{2\rho_{\rm DM}}}{2m_L}\right)
  \simeq
  \frac{\gH^2B_T^2L^2\gL^2\rho_{\rm DM}}{32m_L^2},
  \label{eq:dm-first-band}
\end{equation}
where the last step uses $J_1(x)\simeq x/2$ for a small polarization
rotation amplitude, $x\equiv\gL\sqrt{2\rho_{\rm DM}}/(2m_L)\ll1$.
For comparison, ordinary conversion with $\gL=0$ oscillates according to
Eq.~\eqref{eq:uniform-coherence}. Averaging over these rapid conversion
oscillations replaces $\sin^2(\Delta_{a_H\gamma}L/2)$ by $1/2$,
giving half the upper envelope.
At the first resonance, $\Delta_{a_H\gamma}=m_L$, the enhancement over
this cycle-averaged baseline is
\begin{equation}
  \frac{P^{\rm res}_{a_H \to \gamma}}
  {P^{\rm no \; res}_{a_H \to \gamma}}
  \simeq\frac{\gL^2\rho_{\rm DM} L^2}{16},
  \label{eq:dm-enhancement}
\end{equation}
where $P^{\rm no \; res}_{a_H \to \gamma} \simeq \frac{1}{2}
(\gH B_T/\Delta_{a_H\gamma})^2$ denotes the cycle-averaged
nonresonant probability.
This gain comes from recovering coherent buildup over $L$ in place
of the ordinary fast-oscillation scale $\Delta_{a_H\gamma}^{-1}$.


For a local M82 illustration, we use the following benchmark values given in
Table~\ref{tab:local-m82-benchmark}, based on the model of
Section~\ref{sec:m82-model}.
\begin{table}[htbp]
  \centering
  \begin{tabular}{@{}ll@{}}
    \toprule
    Quantity & Benchmark value\\
    \midrule
    Transverse magnetic field, $B_T$ & $\mathcal O(10)\,\mu{\rm G}$\\
    Light-axion density, $\rho_{\rm DM}$ & $5\,{\rm GeV\,cm^{-3}}$\\
    DM velocity dispersion & $\sim100\,{\rm km\,s^{-1}}$\\
    Light-axion mass, $m_L$ & $10^{-22}\,{\rm eV}$\\
    Light-axion photon coupling, $\gL$ &
      $6\times10^{-13}\,{\rm GeV}^{-1}$\\
    Heavy-axion mass, $m_H$ & $3\times10^{-9}\,{\rm eV}$\\
    Heavy-axion photon coupling, $\gH$ &
      $10^{-12}\,{\rm GeV}^{-1}$\\
    \bottomrule
  \end{tabular}
  \caption{Illustrative inputs for the local DM-induced AMR estimate.
  The full M82 calculation includes spatially varying magnetic field,
  DM density, electron density, and velocity dispersion.
  The $20\%$ light-axion fraction is a benchmark motivated by Lyman-$\alpha$~\cite{Kobayashi:2017ULDM,Liu:2026MixedFDM} and
  rotation-curve~\cite{Bar:2021SPARC} studies.
  }
  \label{tab:local-m82-benchmark}
\end{table}

The first
resonance is near $\omega=m_H^2/2 m_L \simeq 50\,{\rm keV}$. 
For $B_T=10\,\mu{\rm G}$ and a coherent path of $L=100\,{\rm pc}$,
the resonant probability is about $4\times10^{-10}$, compared with a
cycle-averaged nonresonant probability of $2\times10^{-12}$.
The enhancement is therefore about $200$.
Of course, this is only for demonstration purpose to show the strength of the enhancement. 
Computing the full M82 signal also requires averaging over magnetic and DM domains and integrating over the energy spectrum. The M82 model and signal calculation are presented in Section~\ref{sec:m82} with more sophisticated modeling of the environment. 

\begin{figure}[htbp]
  \centering
  \includegraphics[width=0.82\textwidth]{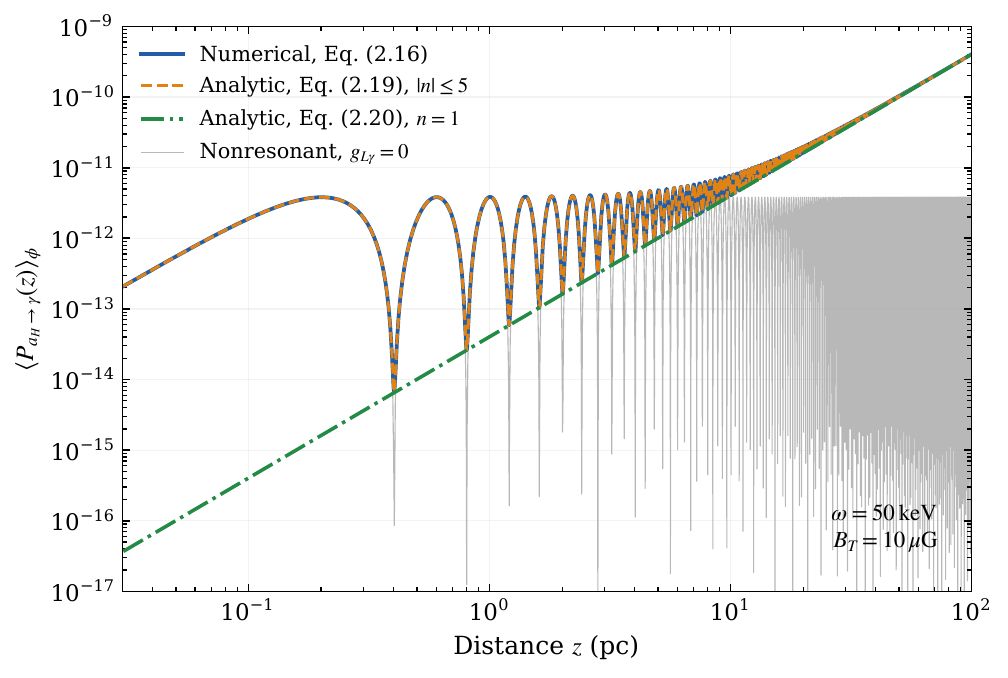}
  \caption{Resonant buildup at $\omega=50\,{\rm keV}$.
  The numerical solution is averaged over 16 uniformly spaced initial
  phases, each held fixed along the path, and compared with
  Eq.~\eqref{eq:dm-harmonic-probability} for $|n|\leq5$ and the
  resonant approximation in Eq.~\eqref{eq:dm-first-band}, which uses
  $J_1(x)\simeq x/2$.
  The thin gray curve shows nonresonant conversion with $\gL=0$.}
  \label{fig:resonance-buildup}
\end{figure}

\begin{figure}[htbp]
  \centering
  \includegraphics[width=0.82\textwidth]{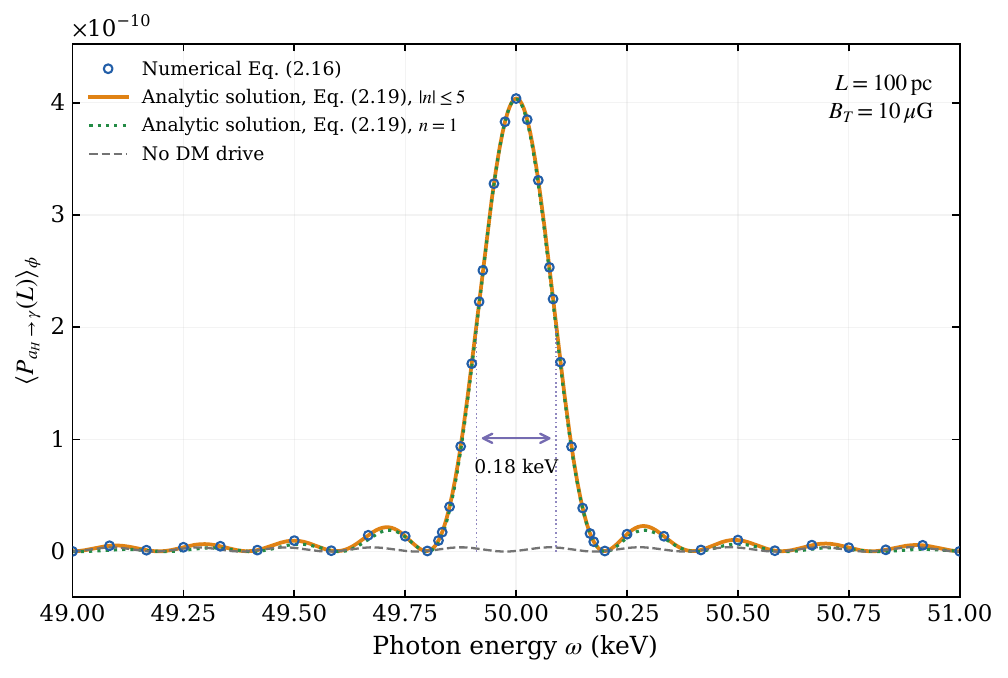}
  \caption{Energy profile of the first resonance for $L=100\,{\rm pc}$.
  The $n=1$ curve uses Eq.~\eqref{eq:dm-harmonic-probability} with
  $J_1(x)\simeq x/2$. The total phase-averaged peak has a full width at
  half maximum of $0.18\,{\rm keV}$; the undriven result is shown for comparison.}
  \label{fig:resonance-width}
\end{figure}

Figures~\ref{fig:resonance-buildup} and~\ref{fig:resonance-width}
illustrate this benchmark with $B_T=10\,\mu{\rm G}$,
$L=100\,{\rm pc}$, and $m_H\simeq3.1623\times10^{-9}\,{\rm eV}$,
chosen to place the first resonance at $50\,{\rm keV}$.
The nonresonant curves retain their rapid conversion oscillations;
Eq.~\eqref{eq:dm-enhancement} instead uses their cycle average.
Plasma dispersion and DM velocities will be added to the full computation in later sections.

\paragraph{Comparison with helical magnetic-field AMR.}
Lastly, before we conclude this section, let us comment on the two types of axion-magnetic resonances, the one induced by a helical magnetic profile and the one due to an oscillating axion DM background. 

Both mechanisms depend on the relative rotation of the photon
polarization vector with respect to the transverse magnetic field. In
the helical magnet induced AMR (Section~\ref{sec:helical-field}), the magnetic field direction rotates;
while here the light axion background rotates the photon polarization.
A medium producing circular birefringence can in principle supply
the same relative rotation and phase matching.

The equivalence can be made more explicit upon applying the inverse transformation that takes 
Eq.~\eqref{eq:rotating-photon-basis} to
Eq.~\eqref{eq:dm-background-propagation}. This removes the
background-induced photon mixing and gives
Eq.~\eqref{eq:helical-laboratory-hamiltonian} with the effective
field $B_T(\sin\theta,\cos\theta,0)$. In this co-rotating basis,
the background-induced rotation is transferred to the field direction.
In principle, this representation holds for any background profile $a_L(z)$ along
the ray. 
Specifically, the oscillations in the background field sets an effective magnetic profile and produce the harmonic resonances derived above.





\section{Application to M82}
\label{sec:m82}

\subsection{Propagation: Resonant X-ray signal}
\label{sec:m82-signal}

\begin{figure}[htbp]
  \centering
  \includegraphics[width=0.7\linewidth]{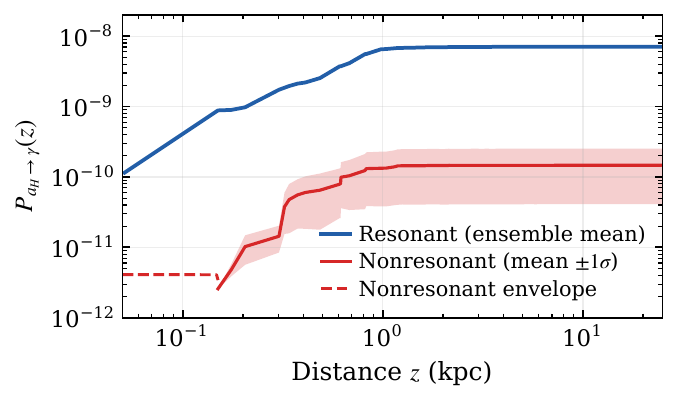}
  \caption{Axion--photon conversion through M82 at $\omega=50\,{\rm keV}$.
  Blue shows the resonant conversion probability averaged over the domain realizations.
  Solid red and shading show the nonresonant magnetic-orientation mean and
  $\pm1\sigma$ realization spread; dashed red traces the upper oscillation
  envelope in the first magnetic domain. In this section, we will mainly explain how we computed the blue curve and focus on quantifying the uncertainty budget for the blue curve.}
  \label{fig:m82-propagation-all-effects}
\end{figure}

Figure~\ref{fig:m82-propagation-all-effects} shows the conversion probability
through M82 at $\omega=50\,{\rm keV}$, comparing resonant and nonresonant
conversion. To obtain the resonant curve, we extend the coherent-region
estimate in Section~\ref{sec:dm-background} to the M82 environment.
That estimate assumes a fixed magnetic direction and DM oscillation phase
over the conversion length. In galaxies such as M82, several effects can affect the resonant
peak height and potentially move conversion power to different energies. We focus on these five effects: 
\begin{itemize}
  \item \textit{Magnetic domains.} Changes of transverse direction of the magnetic field limit
  constructive addition between different parts of the ray. The field
  magnitude also changes the local conversion strength.
  \item \textit{DM domains.} Finite coherence of the axion background limits
  the phase correlation of the induced polarization rotation, introducing
  an additional domain scale in the corotating basis.
  \item \textit{DM density morphology.} The galaxy density profile and
  density fluctuation that peaks at the de Broglie wave length scale change the rotation amplitude, therefore, the resonance strength. 
  \item \textit{Electron density.} Plasma dispersion contributes to the
  photon--axion momentum mismatch and can shift the resonance energy.
  \item \textit{Velocity broadening.} The DM velocity dispersion produces a
  Doppler spreading of driving frequencies along the ray and, hence, that of the resonance
  energy at fixed $m_L$ and $m_H$.
\end{itemize}
In what follows, we will quantify these effects one by one. In the appendix we also put on some stress test on these assumptions to show how much they affect the final results. 
\begin{figure}[htbp]
  \centering
  \includegraphics[width=0.9\textwidth]{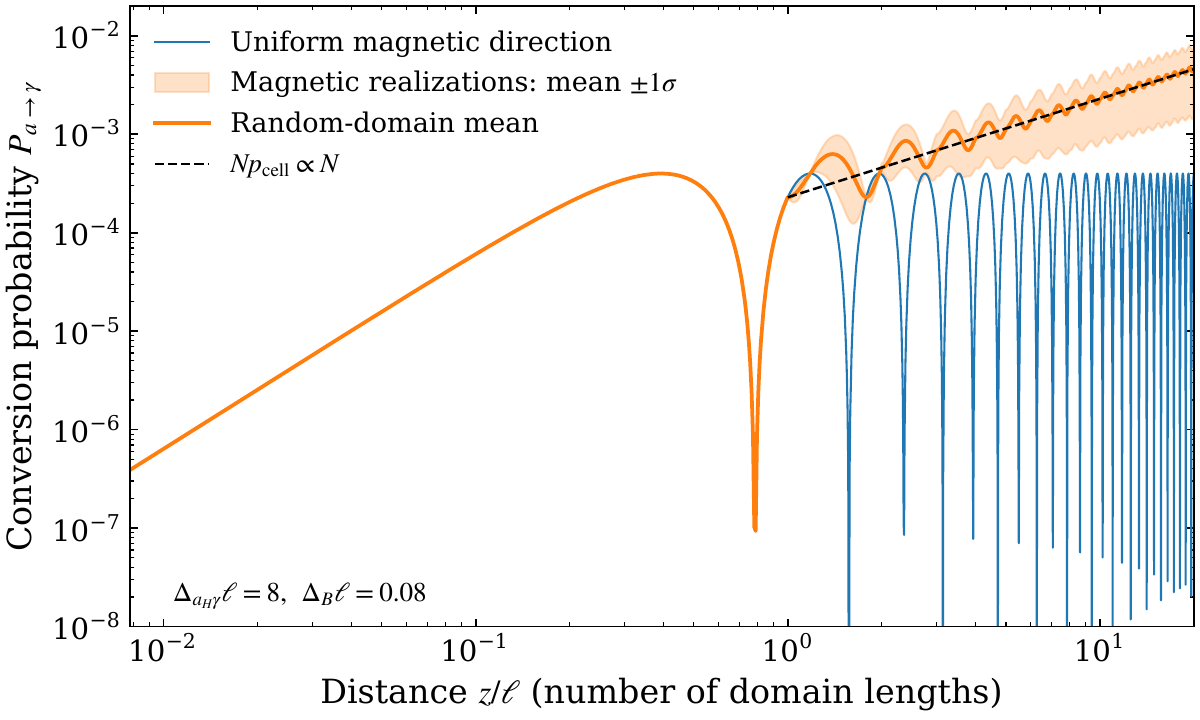}
  \caption{Demonstration of conversion probability enhanced by magnetic domains. 
  Here $\Delta_B=\gH B_\perp/2$ and $\ell$ is the domain length.
 The shaded band shows the mean
  plus or minus one standard deviation across 8192 realizations.
  The dashed line is $N P_{\rm cell}\propto N$, where $P_{\rm cell}$ is
  the single-domain probability, $N$ the number of domains. These illustrative parameters isolate
  magnetic direction changes, without a DM background or M82 profiles. This is a \textit{non-resonant enhancement} purely due to the random orientation of the magnetic field in each domain. }
  \label{fig:magnetic-domain-effect}
\end{figure}
\paragraph{Magnetic domains.}
In M82, the magnetic field is not expected to maintain a single coherent direction
across the whole galaxy. Turbulent gas motions, driven by supernovae and
starburst outflows, stretch and tangle the field lines~\cite{Lacki:2013Starburst}.
An ordered component can coexist with this turbulence, as inferred for
M82~\cite{LopezRodriguez:2021M82}. We approximate the finite magnetic
coherence with a random-cell model that was applied to intergalactic medium (IGM)~\cite{Csaki:2002Dimming,Mirizzi:2009Domains}, galaxy clusters~\cite{Reynolds:2020NGC1275},
as well as galaxies~\cite{Day:2016Galaxies,Carenza:2021Turbulent}. 
In this model, the field direction is assumed to be coherent within each domain and uncorrelated
between domains.

To calibrate the random cell model, we use the example magnetic-field magnitude profile, obtained from an M82 analogue in
the IllustrisTNG TNG50 simulations~\cite{Pillepich:2019TNG50,Nelson:2019TNG50,Ning:2024M82}, as an estimator for the magnetic domain size.
We identify the magnitude plateaus with magnetic domains. 
This gives $\mathcal O(50)$ domains along the 25-kpc ray, while the magnitude $|B(z)|$
follows the interpolated profile. 

Only the field component perpendicular to the propagation direction
contributes to axion--photon mixing. We approximate the projection of an
isotropically oriented field by adopting its RMS transverse amplitude,
$B_\perp(z)=\sqrt{2/3}\,B(z)$, and assign each domain an independent angle
$\psi_i\sim\mathcal U(0,2\pi)$ in the transverse plane, so that
\begin{equation}
  \boldsymbol B_\perp(z)=\sqrt{\frac{2}{3}}\,B(z)
  \bigl(\cos\psi_i,\sin\psi_i\bigr),\qquad z\in\text{domain }i.
  \label{eq:m82-magnetic-domains}
\end{equation}

We highlight the opposite effects of magnetic domains in two regimes,
depending on whether conversion within each domain is in the linear or
nonlinear regime. Here these terms refer to the dependence of the
amplitude on propagation distance: linear buildup or oscillatory behavior.
Consider $N$ equal domains of length $\ell=L/N$, with constant $B_\perp$
and mismatch $\Delta_{a_H\gamma}$. At leading order in $\gH$, independent directions make
the domain probabilities add, giving the limiting scalings
\begin{equation}
  \left\langle P_{a_H\to\gamma}(L)\right\rangle
  \sim \Delta_B^2
  \begin{cases}
    \dfrac{L^2}{N}, & |\Delta_{a_H\gamma}|\ell\ll1
      \quad\text{(small domain size)},\\[6pt]
    \dfrac{N}{\Delta_{a_H\gamma}^2}, & |\Delta_{a_H\gamma}|\ell\gg1
      \quad\text{(large domain size)}.
  \end{cases}
  \label{eq:magnetic-domain-average}
\end{equation}
The second line denotes the oscillation-averaged scaling; both limits
assume small total conversion. In the limit where the domain size is already small, splitting a
coherent path into $N$ domains suppresses its $L^2$ probability by $N$.
In the large domain limit, each domain's contribution already plateaus at the order of
$(\Delta_B/\Delta_{a_H\gamma})^2$, so the total grows with $N$.
Large domain scenario can therefore enhance conversion on
sufficiently long paths, as illustrated in
Fig.~\ref{fig:magnetic-domain-effect}~\cite{Avgoustidis:2010Opacity,BuenAbad:2022Distances}.

\paragraph{Dark matter coherence length.}
DM-assisted conversion depends on the coherence of both the magnetic
field and the axion background. The background rotates photon
polarization by an angle set by Eq.~\eqref{eq:dm-rotation-angle}. At the leading order in $g_{L\gamma}$, and neglecting the plasma effect, the birefringence angle is topological, \textit{i.e.} it only depends on the initial and final axion field value.
In a corotating photon basis, this appears as the evolution of the effective
magnetic-field orientation angle. 
Loss of DM phase correlation therefore
introduces an additional coherence scale even within a single magnetic
domain: an $\mathcal O(1)$ change in the background value $a_L$ leads to an $\mathcal O(g_{L\gamma} a_L)$ change in the magnetic field's effective direction in the corotating frame. 

We approximate this finite coherence by dividing the ray into
independently phased DM domains. A one-dimensional velocity dispersion
$\sigma_{1d}$ gives a momentum spread $m_L\sigma_{1d}$, so we adopt the
reduced de Broglie coherence length. Using the local dispersion from the
isotropic Jeans model, we set
\begin{equation}
  \ell_{\rm DM}(z)=\frac{1}{m_L\sigma_{1d}(z)},
  \qquad
  \int_{z_i}^{z_{i+1}}\frac{dz}{\ell_{\rm DM}(z)}=1,
  \label{eq:m82-dm-domains}
\end{equation}
where $z_i$ are the domain boundaries and velocities are in units of $c$.
The mass model of M82 and the calculation of $\sigma_{1d}(r)$ are given in
Appendix~\ref{app:velocity-dispersion}.
Figure~\ref{fig:m82-dm-domain-size} shows the corresponding local DM
coherence length for the benchmark light-axion mass.

\begin{figure}[htbp]
  \centering
  \includegraphics[width=0.76\textwidth]{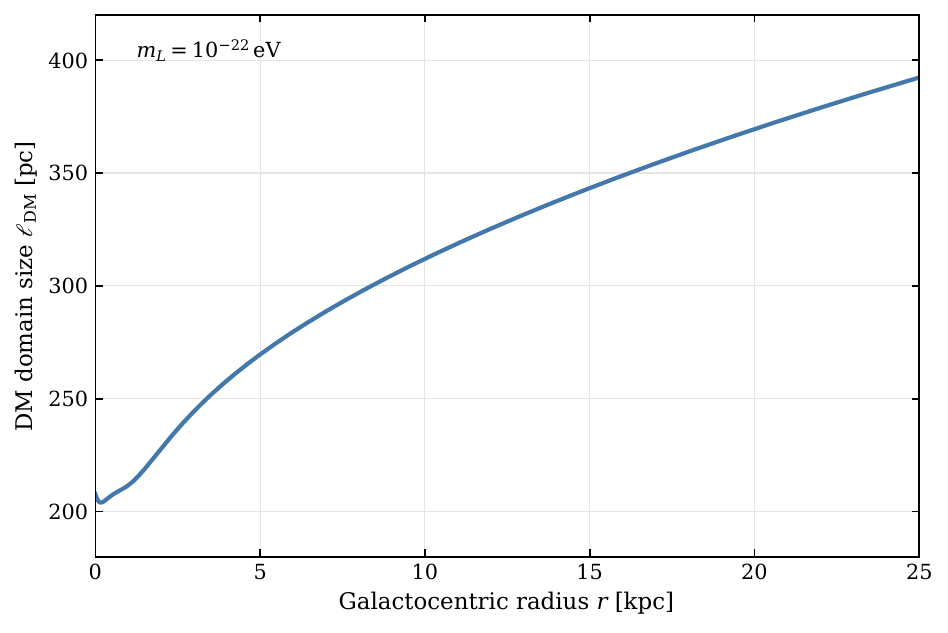}
  \caption{Local DM coherence length in M82 for
  $m_L=10^{-22}\,{\rm eV}$, using the adopted Jeans dispersion and the
  reduced de Broglie convention $\ell_{\rm DM}=1/(m_L\sigma_{1d})$.
  The length scales as $m_L^{-1}$; domain boundaries follow the integral
  condition in Eq.~\eqref{eq:m82-dm-domains}.}
  \label{fig:m82-dm-domain-size}
\end{figure}

Within DM domain $i$, we draw one phase
$\phi_i\sim\mathcal U(0,2\pi)$, independently between domains and from the
magnetic-domain angles. Substituting Eq.~\eqref{eq:dm-background-field}
into Eq.~\eqref{eq:dm-rotation-angle} gives
\begin{equation}
  \theta_i(z)=\frac{\gL}{2}a_{L,i}(z)
  =\frac{\gL\sqrt{2\rho_{\rm DM}(z)}}{2m_L}
    \cos(m_Lz+\phi_i),\qquad z_i\leq z<z_{i+1}.
  \label{eq:m82-dm-domain-angle}
\end{equation}
We omit one common constant angle fixing the photon basis.
If $\psi_j(z)$ is the physical field angle in the magnetic domain containing
$z$, the effective orientation in the corotating basis is
\begin{equation}
  \psi_{\rm eff}(z)=\psi_j(z)-\theta_i(z).
  \label{eq:m82-effective-field-angle}
\end{equation}
 Note that at location $z$, the expression above denotes that it belongs to the $j$-th domain of the magnetic field and the $i$-th domain of the dark matter field. 
At each boundary, we keep the corotating photon amplitudes continuous
and update the effective magnetic direction using
Eq.~\eqref{eq:m82-effective-field-angle}. We add conversion amplitudes along each
realization before averaging probabilities over the random phases and
magnetic directions.

Appendix~\ref{app:m82-wave-packets} examines the spread of
interference-amplitude feature sizes in continuous Gaussian wave fields.
The conversion calculation here retains the domains defined in
Eq.~\eqref{eq:m82-dm-domains}.

\begin{samepage}
\paragraph{DM density morphology.}
We adopt a spherical, smooth core-NFW halo~\cite{Read:2016CoreNFW}
with the M82 parameters of Wang et al.~\cite{Wang:2024M82}.
Its total DM density, $\rho_{\rm halo}(r)$, is finite at the center
and approaches NFW outside the core:
\begin{equation}
  \rho_{\rm halo}(r)=\rho_s\left[
    \frac{\tanh(r/r_c)}{x(1+x)^2}
    +\frac{r_s}{r_c}\frac{f(x)}{x^2}
      \operatorname{sech}^2(r/r_c)
  \right],
  \qquad x=\frac{r}{r_s},
  \label{eq:m82-dm-density}
\end{equation}
\end{samepage}
where $f(x)=\ln(1+x)-x/(1+x)$.
The core and scale radii are $r_c=0.5\,{\rm kpc}$ and
$r_s=3.416\,{\rm kpc}$. The normalization $\rho_s$ gives a halo mass
of $6.0\times10^{10}M_\odot$ within $53\,{\rm kpc}$
(Appendix~\ref{app:m82-mass-model}).

\begin{figure}[htbp]
  \centering
  \includegraphics[width=0.76\textwidth]{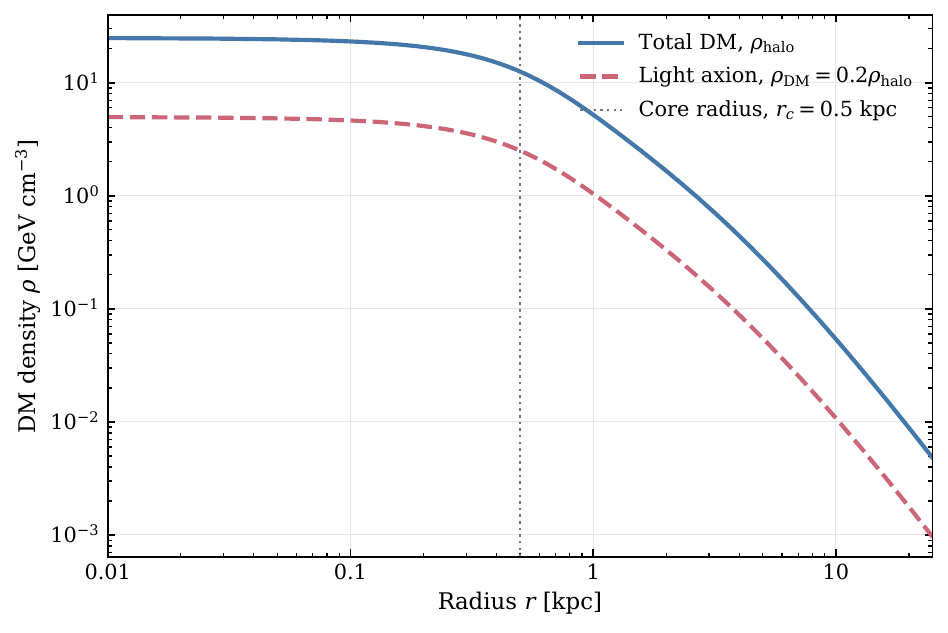}
  \caption{Adopted M82 DM density from Eq.~\eqref{eq:m82-dm-density},
  using the smooth core-NFW prescription~\cite{Read:2016CoreNFW} and
  halo parameters of Wang et al.~\cite{Wang:2024M82}.
  The solid curve shows the total DM density; the dashed curve shows
  the light-axion component, $\rho_{\rm DM}=0.2\rho_{\rm halo}$.
  The vertical dotted line marks the core radius $r_c=0.5\,{\rm kpc}$.
  The finite central densities are approximately $25$ and
  $5\,{\rm GeV\,cm^{-3}}$, respectively. The radial range extends to
  the $25\,{\rm kpc}$ propagation boundary.}
  \label{fig:m82-dm-density}
\end{figure}

The light axion follows the same profile and accounts for only a fraction of
the DM, $\rho_{\rm DM}=0.2\rho_{\rm halo}$, in view of the $\mathcal O(10\%)$ constraints from rotation curves~\cite{Bar:2021SPARC} and Lyman-$\alpha$ bounds~\cite{Kobayashi:2017ULDM,Liu:2026MixedFDM}. The density profile has a central density of about
$5\,{\rm GeV\,cm^{-3}}$. Figure~\ref{fig:m82-dm-density} shows both
density profiles.

The adopted magnetic field is strongest in the inner few
kiloparsecs~\cite{Ning:2024M82}, so this region generally dominates
conversion and makes the resonant signal particularly sensitive to
the inner DM density. In the small-rotation regime,
Eq.~\eqref{eq:dm-first-band} gives
$P_{\rm peak}\propto B_\perp^2\rho_{\rm DM} L^2$ at fixed masses, couplings,
and domain length. A higher density therefore strengthens the
DM-induced polarization rotation and raises the first resonance peak.

\begin{figure}[htbp]
  \centering
  \includegraphics[width=\linewidth]{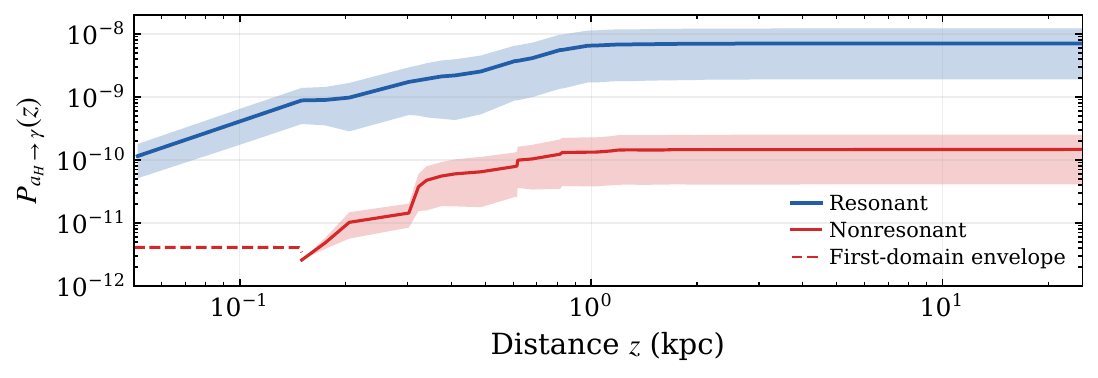}\par\medskip
  \includegraphics[width=\linewidth]{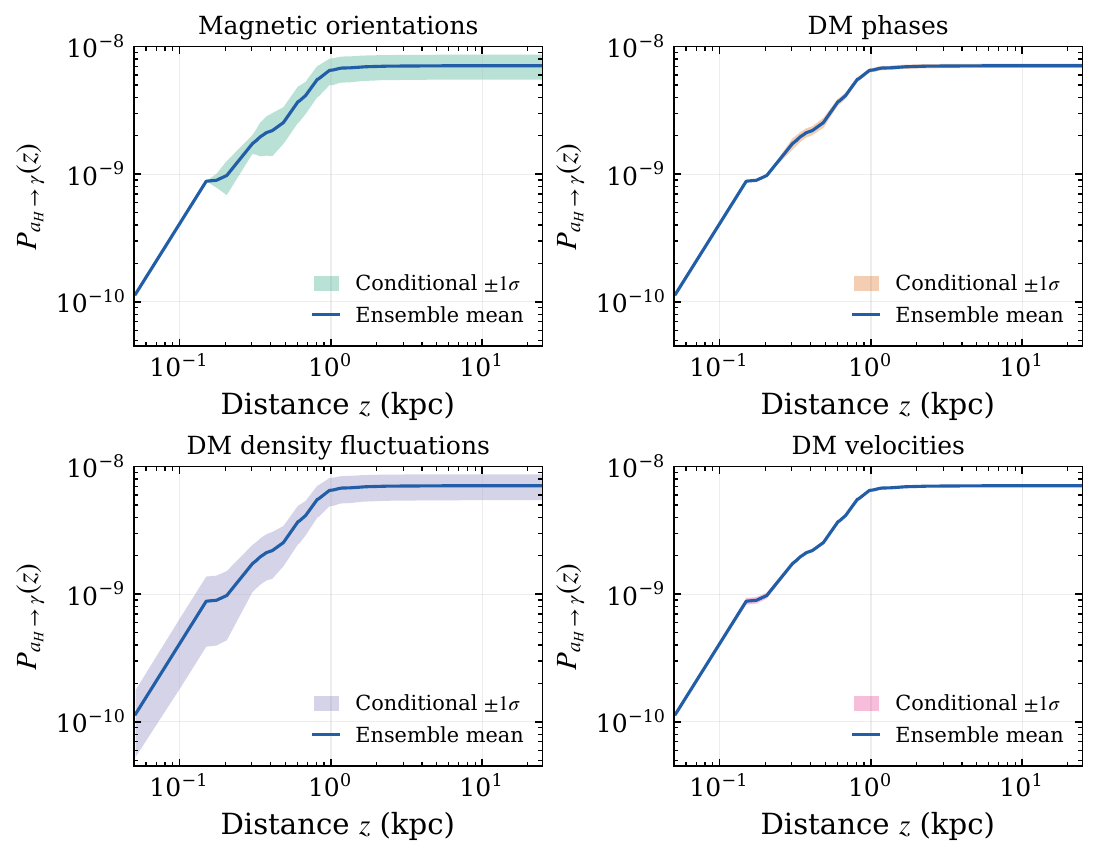}
  \caption{Conversion uncertainty in M82 at $\omega=50\,{\rm keV}$, $m_L=10^{-22}\,{\rm eV}$ and $m_H\simeq3.162\times10^{-9}\,{\rm eV}$. The top panel shows the joint resonant mean and $\pm1\sigma$ realization spread (blue), together with the nonresonant mean (solid red) and $\pm1\sigma$ spread over random magnetic orientations (pink shading). The dashed red curve shows the upper envelope of the rapid nonresonant oscillations in the first magnetic domain, as in Fig.~\ref{fig:m82-propagation-all-effects}. The lower four panels show the common resonant mean and $\pm1\sigma$ of the conditional mean, averaging the other three components. This averaging partially suppresses interference and can narrow the bands relative to tests with the other inputs held fixed. DM-domain sizes remain fixed. Each lower-panel spread uses 256 retained realizations and 128 joint nuisance draws, with phase and magnetic-azimuth nuisance averages performed analytically. The common resonant mean and the top-panel resonant spread use 8,192 joint realizations.}
  \label{fig:m82-resonant-uncertainties}
\end{figure}

The smooth profile describes the mean density. Interference of
light-axion waves can produce $\mathcal O(1)$ density fluctuations on
the DM coherence scale~\cite{Hui:2017ULDM}. These change the local
rotation amplitude as $\sqrt{\rho_{\rm DM}}$, altering peak heights and the
interference pattern along an individual ray. For small rotation
angles, the phase-averaged first-harmonic probability is linear in
density. Independent domain fluctuations with the same mean density
therefore preserve the ensemble-averaged first resonance at leading
order, while adding scatter between realizations. At larger rotation amplitudes, the
nonlinear Bessel weights in Eq.~\eqref{eq:dm-harmonic-probability} can
also change the mean signal and redistribute power among harmonics.
Our fits use the smooth density profile. To test $\mathcal O(1)$ density
fluctuations, we multiply the density in each DM domain by an independent
factor drawn from $\mathcal U(0,2)$. For the $50\,{\rm keV}$ benchmark,
these fluctuations give a $1\sigma$ spread of about $23\%$ in the final
conversion probability at $z=25\,{\rm kpc}$, after averaging over magnetic
orientations, DM phases, and velocities
(Fig.~\ref{fig:m82-resonant-uncertainties}).

\paragraph{Free-electron density.}
Free electrons modify photon propagation even in an electrically neutral
plasma. The relevant quantity is their number density $n_e$, rather than
a net charge imbalance. We adopt the free-electron density profile from
the same simulated M82 analogue used to model the magnetic
field~\cite{Ning:2024M82}. In natural units, the photon plasma mass is
\begin{equation}
\omega_{\rm pl}(z)
  =\sqrt{\frac{4\pi\alpha_{\rm EM}n_e(z)}{m_e}}
  \simeq 3.71\times10^{-11}
  \sqrt{\frac{n_e(z)}{{\rm cm}^{-3}}}\,{\rm eV},
  \label{eq:m82-plasma-mass}
\end{equation}
where $\alpha_{\rm EM}$ is the fine-structure constant and $m_e$ is
the electron mass. The adopted profile reaches
$n_e\simeq2.1\,{\rm cm}^{-3}$, giving
$\omega_{\rm pl}\lesssim5.4\times10^{-11}\,{\rm eV}$.

Substituting this mass into Eq.~\eqref{eq:momentum-mismatch} and
imposing the first-harmonic condition $\Delta_{a_H\gamma}=m_L$ gives
the local resonance energy
\begin{equation}
  \omega_{\rm res}(z)
  =\frac{m_H^2-\omega_{\rm pl}^2(z)}{2m_L}.
  \label{eq:m82-plasma-resonance}
\end{equation}
For the benchmark in Table~\ref{tab:local-m82-benchmark}, the fractional
shift from the vacuum value is at most
$\omega_{\rm pl}^2/m_H^2\lesssim2.8\times10^{-4}$, or
$0.015\,{\rm keV}$ at $51.2\,{\rm keV}$. This is well below the
NuSTAR energy resolution. At this energy, the additional phase
$\int dz\,\omega_{\rm pl}^2/(2\omega)$ within any overlap of a magnetic
domain and a DM domain is below $0.18$, so plasma inhomogeneities do
not appreciably disrupt resonant buildup. Direct numerical comparison
with $n_e=0$ changes the ensemble-averaged first-harmonic peak height
by less than $0.1\%$. Plasma detuning is therefore negligible for this
benchmark resonance. Nevertheless, our numerical propagation retains $n_e(z)$,
since its effect can become significant for smaller $m_H$.

\begin{samepage}
\paragraph{Velocity broadening of the resonance.}
The relevant DM frequency is the rate at which its phase changes
along the axion--photon ray. For a mode with phase
$\omega_Lt-\mathbf k_L\cdot\mathbf x+\phi$ and energy
$\omega_L=\sqrt{m_L^2+|\mathbf k_L|^2}$, following the ray
$t=t_0+z$, $\mathbf x=\mathbf x_0+z\hat{\mathbf z}$ gives
\begin{equation}
  \begin{aligned}
    a_L(z)&\propto\cos\!\left[(\omega_L-k_{L,z})z+\phi_0\right],\\
    \omega_L-k_{L,z}&\simeq m_L\left(1-v_z+\frac{v^2}{2}\right).
  \end{aligned}
  \label{eq:dm-ray-frequency}
\end{equation}
\end{samepage}
Here $\mathbf v=\mathbf k_L/\omega_L$ is in units of $c$, and $\phi_0$ is
a constant phase. The spatial Doppler term $-m_Lv_z$ gives the leading
frequency spread; the kinetic-energy correction enters at order $v^2$.

For a propagating axion/photon of energy $\omega$, the first-harmonic
resonance therefore obeys
\begin{equation}
  \begin{aligned}
    \omega_L-k_{L,z}&=\Delta_{a_H\gamma}
      =\frac{m_H^2-\omega_{\rm pl}^2(z)}{2\omega}
      \simeq\frac{m_H^2}{2\omega},\\
    \omega_{\rm res}(\mathbf v,z)&=\omega_{\rm res}(z)
      \left[1+v_z+\mathcal O(v^2)\right],
  \end{aligned}
  \label{eq:dm-moving-resonance}
\end{equation}
where $\omega_{\rm res}(z)$ is the resonance energy for DM at rest in
Eq.~\eqref{eq:m82-plasma-resonance}; the last expression on the first
line neglects plasma dispersion.

We quantify the impact on conversion by drawing an isotropic Gaussian
velocity in each DM domain, with local component dispersion
$\sigma_{1d}(z)$, and including the frequency shift in
Eq.~\eqref{eq:dm-ray-frequency}. For the $50\,{\rm keV}$ benchmark,
these velocity draws give a $1\sigma$ spread of about $1.8\%$ in the
conversion probability at $z=25\,{\rm kpc}$, after averaging over
magnetic orientations, DM phases, and density fluctuations
(Fig.~\ref{fig:m82-resonant-uncertainties}). The domain boundaries
remain fixed in this comparison.

\paragraph{Uncertainty budget.}
Figure~\ref{fig:m82-resonant-uncertainties} compares the resulting
variations in conversion probability at $\omega=50\,{\rm keV}$, with
the masses and couplings of Section~\ref{sec:dm-background}. We draw
magnetic azimuthal angles, DM phases, density factors
$\delta_{{\rm DM},i}\sim\mathcal U(0,2)$, and Gaussian DM velocities.
The mean density $\bar\rho_{\rm DM}(z)=0.2\rho_{\rm halo}(z)$,
magnetic magnitude $B(z)$, electron density $n_e(z)$, Jeans dispersion
$\sigma_{1d}(z)$, and all domain boundaries remain fixed. For each
realization, we add the conversion amplitudes coherently before
squaring.

The lower four panels isolate each source of variation by varying
that input and averaging over the other three. At $z=25\,{\rm kpc}$,
the fractional $1\sigma$ spreads are $22\%$ for magnetic orientations,
$3.8\%$ for DM phases, $23\%$ for density fluctuations, and $1.8\%$
for velocities, all normalized to the same ensemble mean.
Averaging over magnetic orientations suppresses phase-sensitive
interference, explaining the small conditional DM-phase spread.
When all four inputs vary together, the full spread is $73\%$,
as shown in the top panel. Coherent interference couples their effects,
so the separate conditional spreads do not add in quadrature.

\subsection{Source: axion flux from the stars}
\label{sec:m82-model}

M82 is a starburst galaxy at a distance of approximately
$3.6\,{\rm Mpc}$, with a large population of young, massive stars.
Thermal photons in their hot interiors can convert into axions through
Primakoff scattering off the electric fields of electrons and ions,
as in $\gamma+e^-\to a_H+e^-$.
At the weak couplings considered here, the axions escape freely while
the photons remain trapped. The axion spectrum therefore probes the stellar interior, much hotter than the photosphere. As a result, 
red supergiants, despite their cool surfaces, can be bright sources of
axions with energies of tens of keV~\cite{Ning:2024M82}.

The spectrum of an individual star can be calculated from a model of its
interior at a given age. Utilizing one-dimensional stellar evolution codes, such as MESA (Modules for Experiments in Stellar
Astrophysics), one can solve
the coupled stellar structure and composition equations following
nuclear burning and energy transport~\cite{Paxton:2011MESA,Paxton:2013MESA}.
For each initial mass and composition, it predicts radial profiles of
temperature, density and elemental abundances throughout the star's
life. These profiles determine the thermal photon distribution and the
densities of charged scattering targets; the plasma frequency and
electric screening scale follow from the same local conditions.
Integrating the Primakoff emission over the stellar volume gives the
axion luminosity spectrum at that age.

The galaxy's luminosity is then obtained by summing over stars with
different initial masses and ages. Their relative numbers depend on
the initial mass function and the star-formation history, inferred for
M82 from infrared spectroscopy and population
modeling~\cite{ForsterSchreiber:2003M82}.
The adopted model has a Salpeter-like mass distribution at high masses,
flattening below about $3\,M_\odot$, and two starbursts with inferred ages
$4.1^{+0.5}_{-0.7}$ and $9.0\pm0.2\,{\rm Myr}$.
In the source calculation of Ref.~\cite{Ning:2024M82}, stellar ages are
drawn only from these two burst components, with the population normalized
to M82's estimated total stellar count. No separate population predating
the bursts is included in the spectrum we adopt.
Red supergiants and O-type stars dominate
the resulting axion emission in the $30$--$70\,{\rm keV}$
band~\cite{Ning:2024M82}.

Since the stellar profiles are model predictions, convective mixing and mass loss remain
sources of uncertainty in massive-star evolution~\cite{Paxton:2013MESA}.
The population inference also leaves uncertainties in the abundance and
ages of the stars supplying the hard axion emission.
The source calculation tests a roughly $20\%$ uncertainty in the total
number of stars, a metallicity $Z=0.02\pm0.004$, variations of the burst
parameters, and an increase of the upper initial-mass cutoff from
$100$ to $200\,M_\odot$. These variations change the published low-mass
coupling limits less than the magnetic-field choices do
 (Figure~S20 of Ref.~\cite{Ning:2024M82}). 

We adopt the total M82 axion luminosity spectrum computed in
\cite{Ning:2024M82}. 
Denoting this energy luminosity at
$g_0=10^{-12}\,{\rm GeV}^{-1}$ by $dL_H^{(0)}/d\omega$, the axion number flux
at Earth before conversion is
\begin{equation}
  \frac{d\Phi_H}{d\omega}
  =\frac{1}{4\pi d^2 \omega}
   \left(\frac{\gH}{g_0}\right)^2
   \frac{dL_H^{(0)}}{d\omega},
  \qquad d=3.6\,{\rm Mpc}.
  \label{eq:m82-stellar-axion-flux}
\end{equation}
The factor $1/\omega$ converts energy luminosity into number luminosity,
and $1/(4\pi d^2)$ accounts for geometric dilution.
The quadratic coupling dependence follows from the Primakoff production
rate, neglecting axion cooling effects on stellar evolution and
axion-mass corrections at keV energies.

Lastly, we comment on the uncertainties in terms of the source computation. Since our purpose of this work is to show the resonant enhancement, if a systematic shift in the source affects the constraint contour, it should do so similarly to the cases  with (this work) and without (e.g. \cite{Ning:2024M82}) the resonance. Therefore, it doesn't hinder the main message which is the improvement of the DM-induced resonance conversion.

\subsection{Detector: NuSTAR X-ray observations}
\label{sec:ma-ga}

To compare a photon-flux model with the NuSTAR~\cite{Harrison:2013NuSTAR} net counts
constructed in Section~\ref{sec:m82-data}, we account for the
detector's collection efficiency through the effective area
$A_{\rm eff}(\omega)$ and its imperfect energy measurement through $R_i(\omega)$,
the probability that a photon of true energy $\omega$ is registered in bin $i$.
The predicted net counts are
\begin{equation}
  \mu_i^{\rm th}=T_{\rm tot}\int d\omega\,A_{\rm eff}(\omega)R_i(\omega)
  \frac{d\Phi_\gamma}{d\omega}.
  \label{eq:nustar-folded-counts}
\end{equation}
Here $d\Phi_\gamma/d\omega$ denotes the total modeled photon flux contributing
to the source excess after subtracting the surrounding background.
For the M82 fit, it comprises converted axion photons and the residual astrophysical
continuum, which we describe phenomenologically by a power
law~\cite{Ning:2024M82}:
\begin{equation}
  \frac{d\Phi_\gamma}{d\omega}
  =\frac{d\Phi_{a\to\gamma}}{d\omega}
   +\frac{d\Phi_{\rm astro}}{d\omega},\qquad
  \frac{d\Phi_{\rm astro}}{d\omega}
  =A\left(\frac{\omega}{50\,{\rm keV}}\right)^{-\Gamma}.
  \label{eq:nustar-photon-components}
\end{equation}
The normalization $A\geq0$ and index $\Gamma$ are nuisance parameters;
both photon components are passed through the same adopted response.

We use the one-module effective-area curve from ObsID~30202022004, together with
the mission's representative near-axis redistribution
matrix~\cite{NuSTAR:Responses}. Its $4096$ true-energy intervals span
$1.60$--$165.44\,{\rm keV}$. We sum the redistribution probabilities over
the measured-energy bins
defined in Section~\ref{sec:m82-data}, retaining broadening, escape tails
and probability losses.
The true-energy integral extends approximately
from $28.54$ to $88.67\,{\rm keV}$, including redistribution into the
analysis band from outside it; its upper endpoint is set by the adopted
effective-area table. We use a baseline $0.02\,{\rm keV}$ continuum
grid supplemented by response nodes and points resolving narrow
resonances, with finer-grid checks of the reported fits.
The common effective area and redistribution approximate the responses
of the individual observations.

\subsection{Data: selection and processing}
\label{sec:m82-data}

For a close comparison with the non-resonant conversion computed in \cite{Ning:2024M82}, we use the same NuSTAR observation data selection as theirs.
We independently repeat the extraction and rebinning for the same
37 archival observations, following their source and background selections
and using both NuSTAR focal-plane modules, FPMA and FPMB.
The starting data for this check are public, calibrated and screened
Level-2 event lists. Each event records an arrival time, a reconstructed
sky position and a measured energy channel (pulse invariant, PI).
We retain the archived good-time intervals and extract source and
background spectra with \texttt{nuproducts} in
HEASoft~6.37.1/NuSTARDAS~2.1.6. The apertures are reconstructed from
their published region geometry and listed in
Table~\ref{tab:nustar-settings}.

\begin{table}[tbp]
  \centering
  \small
  \begin{tabular}{p{0.28\textwidth}p{0.64\textwidth}}
    \toprule
    Quantity & Adopted value\\
    \midrule
    Response exposure & $2.072864\,{\rm Ms}$ per module;
      total $T_{\rm tot}=4.145728\,{\rm Ms}$\\
    Aperture center (J2000) &
      RA $148.965^\circ$, Dec $69.679^\circ$\\
    Source region (ON) & Circle of radius $49.2''$
      (nominally $68\%$ PSF containment)\\
    Background region (OFF) & Annulus with radii $81.18''$--$105.78''$
      (between the nominal $80\%$ and $90\%$ encircled-energy radii)\\
    Recorded-energy range & $30\leq \omega_{\rm det}<70\,{\rm keV}$\\
    Native PI channels & $0.04\,{\rm keV}$ wide;
      $\omega_{\rm low}=(1.6+0.04\,{\rm PI})\,{\rm keV}$\\
    Rebinned spectra & Eight $5\,{\rm keV}$ bins for the results below;
      $1\,{\rm keV}$ and $0.4\,{\rm keV}$ bins for processing checks\\
    Nominal spectral resolution & FWHM $0.4\,{\rm keV}$ at
      $10\,{\rm keV}$ and $0.9\,{\rm keV}$ at $68\,{\rm keV}$\\
    \bottomrule
  \end{tabular}
  \caption{NuSTAR selections and energy groupings. For the representative
  response, we assign the summed observation times in Table~S1 of
  Ref.~\cite{Ning:2024M82} to each module. The total includes both modules.
  Native channel spacing and
  physical energy resolution are distinct~\cite{NuSTAR:FAQ,NuSTAR:Technical}.}
  \label{tab:nustar-settings}
\end{table}

We add the source and background counts separately over all observations
and both modules. Rebinning sums whole native channels: $125$ per
$5\,{\rm keV}$ bin, $25$ per $1\,{\rm keV}$ bin, and $10$ per
$0.4\,{\rm keV}$ bin.
Let $S_i$ and $B_i$ be the measured counts in energy bin $i$ in the
source circle and surrounding annulus, respectively. With the geometric
source-to-background area ratio $\alpha=10/19$, the experimental net
counts and their estimated standard deviations are
\begin{equation}
  \mu_i^{\rm exp}=S_i-\alpha B_i,\qquad
  \sigma_i=\sqrt{S_i+\alpha^2B_i}.
  \label{eq:nustar-net-counts}
\end{equation}
The uncertainty follows by adding the independent Poisson variances,
with the unknown means estimated by the observed counts and $\alpha$
held fixed. Counts and variances $\sigma_i^2$ are summed when grouping;
negative net bins are retained.

The background annulus estimates instrumental and diffuse X-ray counts.
Emission concentrated toward M82 can remain after this subtraction,
including hard X-rays from its ultraluminous X-ray
sources~\cite{Brightman:2020M82ULX}. We therefore capture  this remaining emission
by the power law in Eq.~\eqref{eq:nustar-photon-components}.

For validation of our data processing, our independent reduction gives
$2.7\%$ fewer total net counts than the tabulated spectrum of
Ref.~\cite{Ning:2024M82} over $30$--$70\,{\rm keV}$.
The largest relative difference in an individual $5\,{\rm keV}$ bin
is $10.6\%$ of the tabulated net count.
For a close comparison, we use the tabulated data provided in \cite{Ning:2024M82} for the 5 keV fits, while we use our own rebinned data processed directly from NuSTAR release for other fits. 

We compare the measured net counts with the prediction using the
Gaussian count likelihood
\begin{equation}
  -2\ln\mathcal L
  =\sum_i\frac{\bigl(\mu_i^{\rm exp}-\mu_i^{\rm th}\bigr)^2}
                  {\sigma_i^2}+\mathrm{const},
  \label{eq:nustar-likelihood}
\end{equation}
with fixed observed variances and all fitted bins retained.

\section{Results}
\label{sec:results}
\subsection{Constraints}
\label{sec:enhanced-constraints}

\begin{figure}[htbp]
  \centering
  \includegraphics[width=0.75\textwidth]{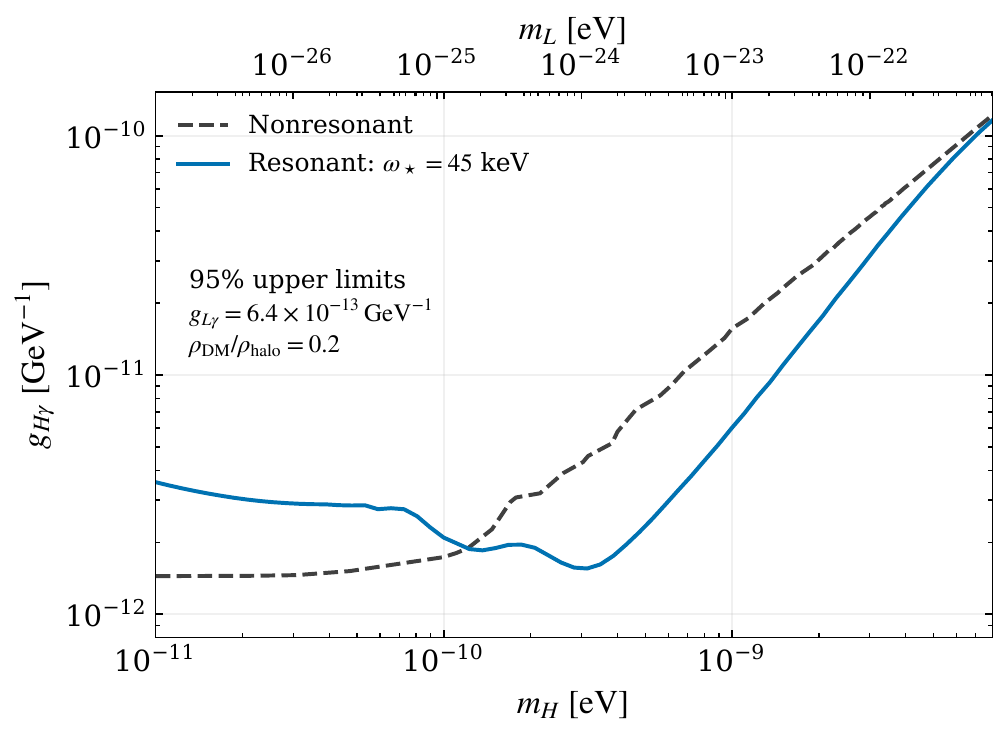}
  \caption{$95\%$ upper limits on $\gH$ with (blue) and without
  (dashed) the DM background. The blue curve fixes
  $\omega_\star=45\,{\rm keV}$; the upper axis shows the corresponding
  $m_L=m_H^2/(2\omega_\star)$.}
  \label{fig:m82-enhanced-constraints}
\end{figure}

In Figure~\ref{fig:m82-enhanced-constraints} we show the $95\%$
upper limits on $\gH$ with and without the DM background.
Throughout this section we fix $\omega_\star=m_H^2/(2m_L)=45\,{\rm keV}$,
so $m_L$ is determined by $m_H$.
We fit the spectrum described in Section~\ref{sec:m82-data}, using the
central $25\,{\rm kpc}$ ray.
For the DM background we fix $\gL=6.4\times10^{-13}\,{\rm GeV}^{-1}$ to stay below the strongest constraint to-date
and $\rho_{\rm DM}(z)=0.2\rho_{\rm halo}(z)$.
Here $\rho_{\rm DM}$ is the light-axion density and $\rho_{\rm halo}$
is the total DM density, following the notation in Section~\ref{sec:m82}.

The predicted signal uses the conversion probability averaged over the
unknown magnetic directions and DM phases. Within each magnetic domain,
the field direction is fixed; we average its angle in the plane
perpendicular to the ray uniformly from $0$ to $2\pi$.
We also average the starting phase of the light-field oscillation
uniformly from $0$ to $2\pi$ in each DM coherence region.
All these angles and phases are independent, and the magnetic and DM
regions have separate boundaries. Masses, couplings and spatial profiles
are held fixed during this averaging.
The DM density follows the smooth profile above, without random density
fluctuations, as we saw that the impact of the fluctuations is small. The DM velocity spread calculated from the Jeans equation
sets the sizes of the coherence regions. We do not include individual velocities and their
Doppler shifts, since their impact is small.

\begin{figure}[htbp]
  \centering
  \includegraphics[width=\textwidth]{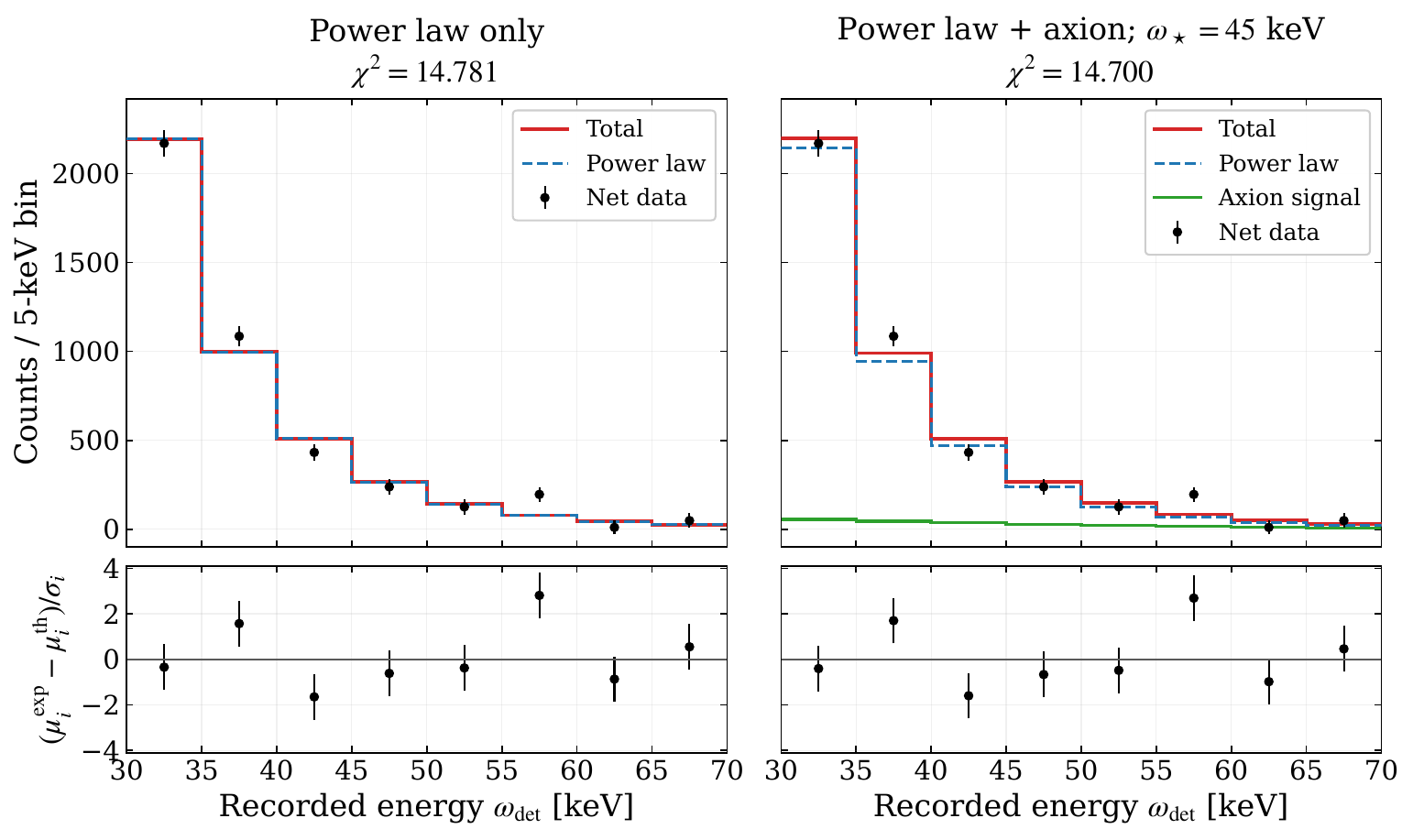}
  \caption{Best fits to the NuSTAR spectrum: a power law alone (left)
  and a power law plus the axion signal at fixed
  $\omega_\star=45\,{\rm keV}$ (right). Lower panels show the normalized
  residuals $(\mu_i^{\rm exp}-\mu_i^{\rm th})/\sigma_i$.}
  \label{fig:m82-fixed45-residuals}
\end{figure}

The axion signal being fitted consists of X-ray photons produced when
stellar $a_H$ axions convert in M82's magnetic field. The light-axion DM
background modifies this conversion probability. At each trial axion
parameter point, the converted photon flux is the stellar axion flux in
Eq.~\eqref{eq:m82-stellar-axion-flux} multiplied by the mean conversion
probability:
\[
  \frac{d\Phi_{a\to\gamma}}{d\omega}
  =\left\langle P_{a_H\to\gamma}(\omega)\right\rangle
   \frac{d\Phi_H}{d\omega}.
\]
Applying the detector response in Eq.~\eqref{eq:nustar-folded-counts}
gives the predicted axion-induced photon counts in each measured bin.
These counts are added to the power-law counts to form the model
compared with the data. This signal calculation is repeated as the axion
parameters vary. The galaxy profiles remain fixed.
We then perform the following fits:
\begin{enumerate}
  \item Fit the power law alone as a reference model, varying its normalization
  $A\geq0$ and slope $\Gamma$ to minimize the $\chi^2$ in
  Eq.~\eqref{eq:nustar-likelihood}, keeping the measured errors fixed.
  \item Fit the power law plus the predicted axion-induced photon counts.
  Vary $m_H,\gH$, setting $m_L=m_H^2/(2\omega_\star)$ with
  $\omega_\star=45\,{\rm keV}$, and refit $A$ and $\Gamma$ at every
  trial. The background parameters are free in this fit; their values
  from the power-law-only fit are not held fixed.
  The lowest $\chi^2$ found in this
  search gives the four-parameter fit shown in
  Figure~\ref{fig:m82-fixed45-residuals}.
  \item For an upper limit, hold the masses fixed and vary $\gH$, again
  refitting $A$ and $\Gamma$ at every trial. The $95\%$ upper limit is
  where $\chi^2$ rises by $2.71$ above its minimum at those masses,
  on the side with larger $\gH$.
\end{enumerate}

We scan $10^{-11}\leq m_H\leq8\times10^{-9}\,{\rm eV}$,
keeping $\omega_\star=45\,{\rm keV}$ fixed for both the upper limits and
the best-fit spectrum.
For the comparison curve, we omit the DM background from the conversion
calculation and repeat the fit.

\paragraph{The goodness of fit.}
Figure~\ref{fig:m82-fixed45-residuals} compares the power law alone with
the power law plus the axion signal at fixed $\omega_\star=45\,{\rm keV}$.
The axion fit varies $m_H,\gH,A,\Gamma$; $m_L$ follows from the fixed
resonance energy. The resonance leads to a deeper reach in the constraint contour at $m_H \gtrsim 10^{-10}\,\mathrm{eV}$. 


\paragraph{Enhancement vs suppression. }The suppression at small $m_H$ follows from the full harmonic sum in
Eq.~\eqref{eq:dm-harmonic-probability}. Consider a domain of length $L$
with constant $B_T$ and $\rho_{\rm DM}$, where
$|\Delta_{a_H\gamma}|L\ll1$. Ordinary conversion is already phase-matched,
with probability $P_{\rm coh}=(\gH B_TL/2)^2$. Writing the rotation
amplitude as $\beta=\gL\sqrt{2\rho_{\rm DM}}/(2m_L)$, we obtain
\begin{equation}
  \frac{\langle P_{a_H\to\gamma}\rangle_\phi}{P_{\rm coh}}
  \simeq\sum_{n=-\infty}^{\infty}J_n^2(\beta)\,
  \operatorname{sinc}^2\!\left[
    \frac{(\Delta_{a_H\gamma}-nm_L)L}{2}
  \right]\leq1 .
  \label{eq:low-mass-harmonic-bound}
\end{equation}
The inequality follows from $\sum_nJ_n^2(\beta)=1$ and
$\operatorname{sinc}^2x\leq1$. The DM background redistributes the
harmonic weights, while the sinc factors reduce contributions that fail
to add constructively. It cannot increase the probability above the
fully phase-matched limit.

As $m_L\to0$, the harmonic mismatch $nm_L$ vanishes at every fixed
$n$. However, the weights
$J_n^2(\beta)$ spread over orders $|n|\sim\beta$ as $\beta$ grows;
their root mean square order is $\beta/\sqrt{2}$. At fixed
$\omega_\star$, lowering $m_H$ lowers $m_L=m_H^2/(2\omega_\star)$
and increases $\beta\propto1/m_L$. Thus the harmonics carrying the
weight move to larger $|n|$, and their phase mismatches remain finite:
\begin{equation}
  |n|m_LL\sim\beta m_LL
  =\frac{\gL\sqrt{2\rho_{\rm DM}}\,L}{2}\equiv D .
  \label{eq:low-mass-harmonic-scale}
\end{equation}
For $D\gg1$, the corresponding sinc factors are small. Taking the limit
at fixed $n$ therefore misses this suppression. The isolated first-band expression in
Eq.~\eqref{eq:dm-first-band}, whose final approximation also assumes
weak rotation, cannot describe this regime.

The limiting result can also be obtained by expanding the slowly
oscillating background: $\theta(z)-\theta(0)\to-(D/L)z\sin\phi$.
Integrating this phase along the domain and averaging over $\phi$ gives
\begin{equation}
  \frac{\langle P_{a_H\to\gamma}\rangle_\phi}{P_{\rm coh}}
  \xrightarrow[\Delta_{a_H\gamma}L\to0]{m_L\to0}
  \frac{1}{2\pi}\int_0^{2\pi}d\phi\,
  \operatorname{sinc}^2\!\left(\frac D2\sin\phi\right)
  \simeq\frac{2}{D}\quad(D\gg1).
  \label{eq:low-mass-suppression}
\end{equation}
At fixed density and domain length, this ratio tends to a constant.
Equation~\eqref{eq:low-mass-suppression} therefore shows why suppression
can persist, but does not explain the continued rise of the upper limit
at the lowest masses in Figure~\ref{fig:m82-enhanced-constraints}.
In the plotted calculation, the density varies within each magnetic
domain. Differentiating Eq.~\eqref{eq:m82-dm-domain-angle} gives
\begin{equation}
  \begin{aligned}
    \frac{d\theta}{dz}
    &={\frac{\gL}{2m_L}}
      \frac{d\sqrt{2\rho_{\rm DM}(z)}}{dz}\cos(m_Lz+\phi)\\
    &\quad-\frac{\gL\sqrt{2\rho_{\rm DM}(z)}}{2}\sin(m_Lz+\phi).
  \end{aligned}
  \label{eq:low-mass-density-rotation}
\end{equation}
The first term vanishes for uniform density but grows as $1/m_L$ for
the adopted spatial profile, producing additional cancellation and the
low-mass upturn in Figure~\ref{fig:m82-enhanced-constraints}.
However, when its characteristic size becomes comparable to the
oscillation term, the field amplitude changes appreciably over a
Compton length $1/m_L$. This violates the assumption of a slowly
varying field amplitude underlying the local oscillating-DM description.
The coherence length $1/(m_L\sigma_{1d})$ is larger still and can exceed
the galaxy size. Coherence across the galaxy is not itself inconsistent,
but imposing the same galactic density profile in this regime requires
a self-consistent wave description~\cite{Hui:2017ULDM}.
The rising tail therefore depends on extrapolating the adopted DM
profile beyond its established validity and should not be interpreted
as a reliable prediction of stronger suppression.

\paragraph{Comparison with existing bounds. } At small $m_H$, our nonresonant upper limit is slightly weaker than that
of Ning and Safdi~\cite{Ning:2024M82}. Possible sources of this difference
include our representative detector response, details of the likelihood
implementation, and the use of a central-ray magnetic-domain model
instead of their source-averaged three-dimensional calculation. We have
not isolated the contribution of each approximation. Here we focus on
the enhancement produced by the light-axion background relative to
ordinary conversion. Both contours use the same data, source spectrum,
detector response, galaxy model, and fitting procedure. Common
normalization shifts therefore largely cancel in their comparison, so a
small offset in the absolute nonresonant limit should not affect the amount of 
resonant improvement.

\FloatBarrier

\subsection{Resonance features}
\label{sec:resonance-features}

Resonant conversion can produce a spectral feature whose energy depends
on the axion masses. We make a tentative search for such features by
allowing the resonance energy $\omega_\star=m_H^2/(2m_L)$ to vary.
The two-axion model has five free parameters,
$m_H,m_L,\gH,A,\Gamma$, with $\omega_\star=m_H^2/(2m_L)$.
We keep $\gL$ and $\rho_{\rm DM}/\rho_{\rm halo}$ fixed as in
Section~\ref{sec:enhanced-constraints}.

We compare this model with a power law alone, which fits $A,\Gamma$,
and a single-axion model with ordinary conversion, which fits
$m_H,\gH,A,\Gamma$. Thus the resonance model adds three parameters
to the power law and one to the single-axion model.
All fits use the same count likelihood and refit the power-law
normalization and slope at each trial.

\begin{figure}[htbp]
  \centering
  \includegraphics[width=\textwidth]{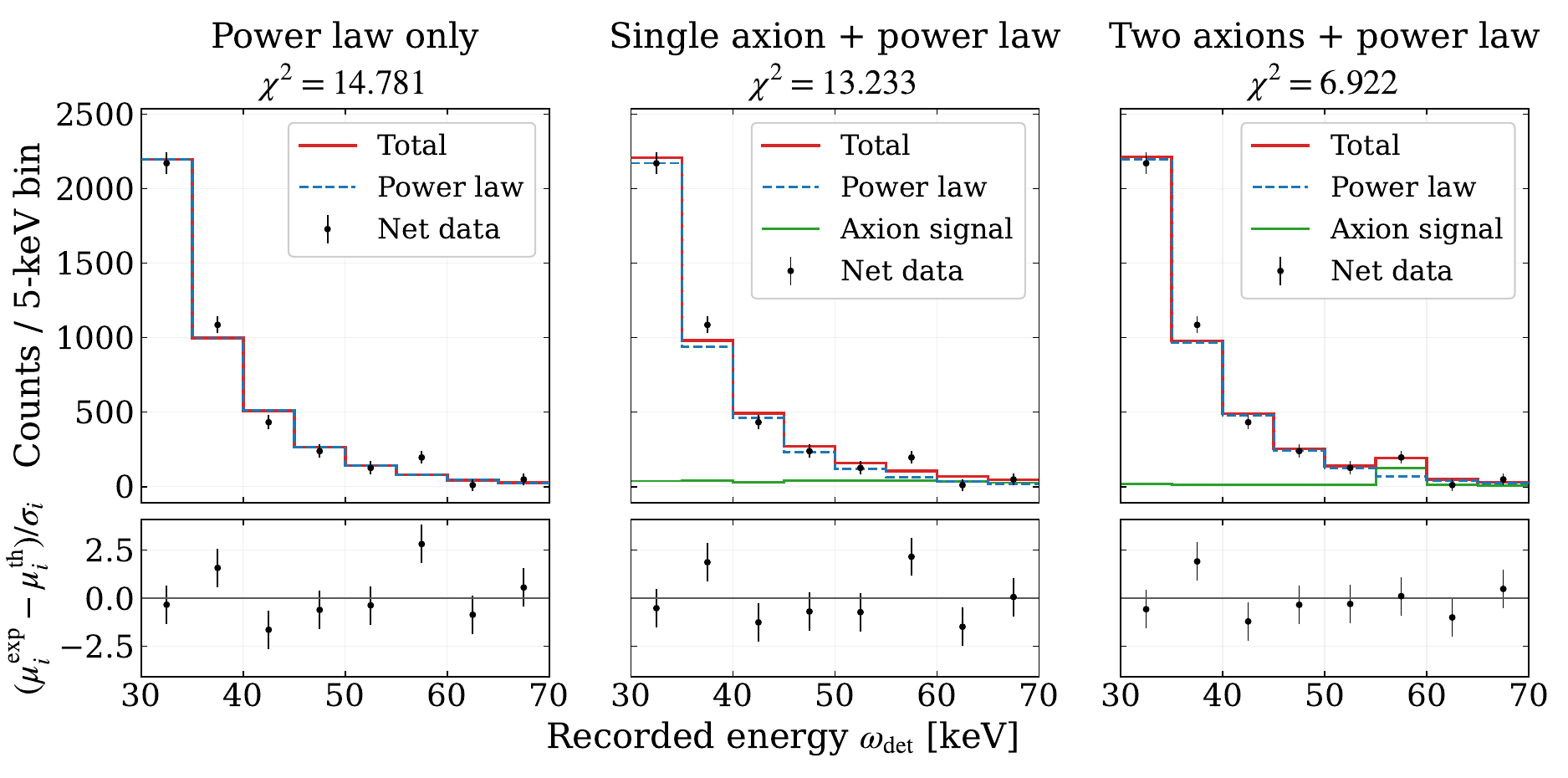}
  \caption{Fits to the $5\,{\rm keV}$ NuSTAR spectrum:
  power law alone (left), single axion plus power law (middle), and
  two axions plus power law (right).
  The single-axion model uses ordinary conversion. In the two-axion model,
  the DM background modifies the conversion of heavy axions produced in
  stars. We fit the resonance energy $\omega_\star$ in the right panel,
  keeping $\gL=6.4\times10^{-13}\,{\rm GeV}^{-1}$ and
  $\rho_{\rm DM}/\rho_{\rm halo}=0.2$ fixed.
  Both power-law parameters are fitted separately in each model.
  Steps show counts after applying the detector response. Lower panels
  show $(\mu_i^{\rm exp}-\mu_i^{\rm th})/\sigma_i$, with the same definitions
  as in Figure~\ref{fig:m82-fixed45-residuals}.}
  \label{fig:m82-floating-5kev}
\end{figure}

Figure~\ref{fig:m82-floating-5kev} shows the three fits to the
$5\,{\rm keV}$ spectrum used in Section~\ref{sec:enhanced-constraints}.
The best two-axion fit found has $\omega_\star=58.4\,{\rm keV}$,
$m_H=1.62\times10^{-9}\,{\rm eV}$,
$m_L=2.25\times10^{-23}\,{\rm eV}$ and
$\gH=1.43\times10^{-11}\,{\rm GeV}^{-1}$.
The single-axion fit has $m_H=6.27\times10^{-10}\,{\rm eV}$ and
$\gH=7.72\times10^{-12}\,{\rm GeV}^{-1}$.
The two-axion fit lowers $\chi^2$ by $7.859$ relative to the power law and by
$6.311$ relative to the single-axion fit.
Table~\ref{tab:m82-floating-5kev} gives the fit quality for all three models.

\begin{table}[htbp]
  \centering
  \small
\begin{tabular}{@{}lrrrrr@{}}
\toprule
Model & $k$ & $\chi^2$ & $\nu$ & $\chi^2/\nu$ & $\Delta\chi^2_{\rm PL}$\\
\midrule
Power law only & 2 & 14.781 & 6 & 2.464 & 0.000\\
Single axion + power law & 4 & 13.233 & 4 & 3.308 & -1.548\\
Two axions + power law & 5 & 6.922 & 3 & 2.307 & -7.859\\
\bottomrule
\end{tabular}
  \caption{Goodness of fit for the eight $5\,{\rm keV}$ bins
  in Figure~\ref{fig:m82-floating-5kev}. Here $k$ is the number of fitted
  parameters and $\nu=8-k$. We define
  $\Delta\chi^2_{\rm PL}=\chi^2_{\rm model}-\chi^2_{\rm PL}$;
  negative values indicate a better fit than the power law alone.
  The listed $\nu$ and reduced $\chi^2$ are for comparison; the significance
  of the improvements is estimated separately.}
  \label{tab:m82-floating-5kev}
\end{table}

\begin{figure}[htbp]
  \centering
  \includegraphics[width=\textwidth]{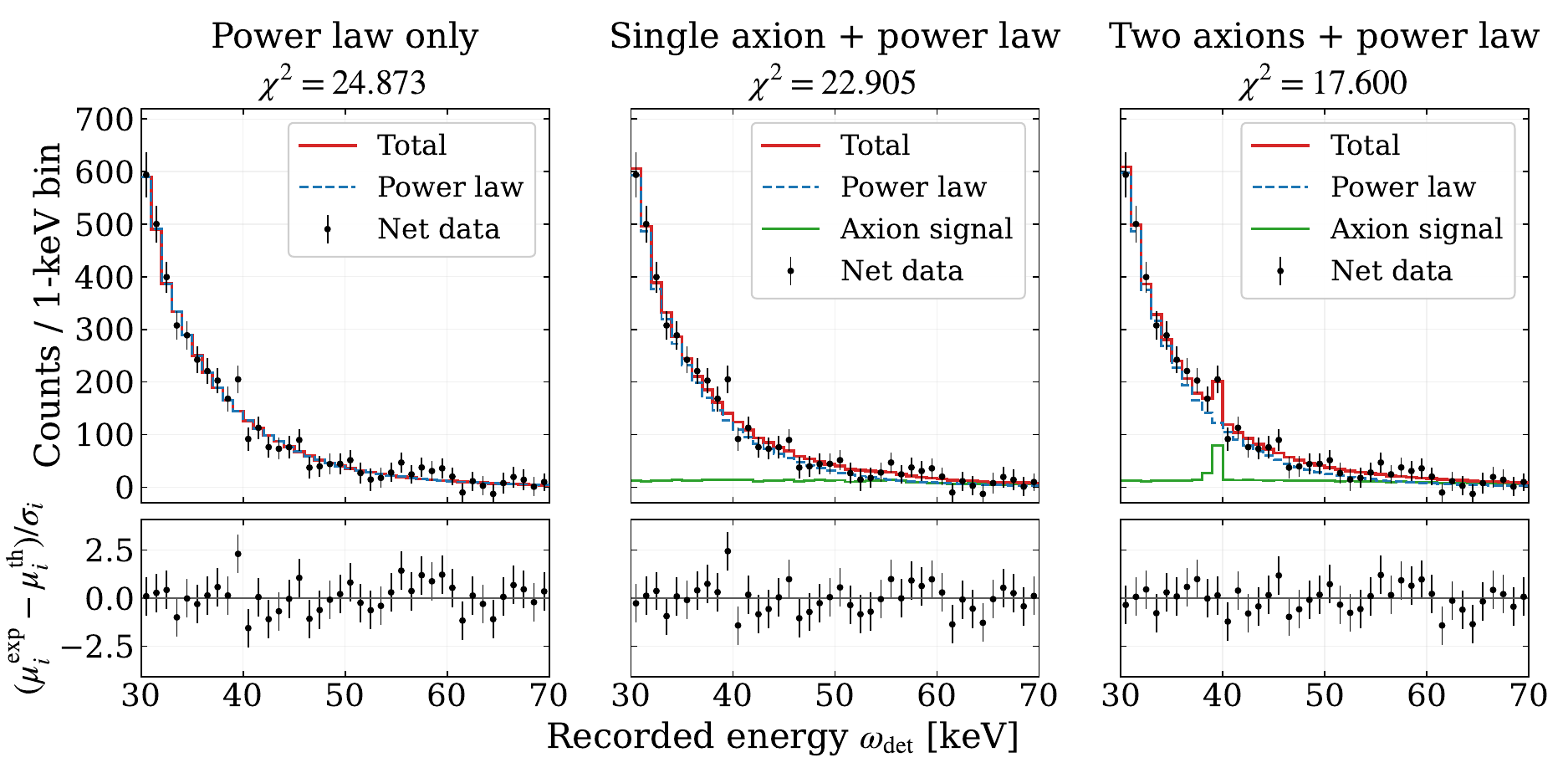}
  \caption{Fits to the separately extracted $1\,{\rm keV}$ NuSTAR spectrum,
  with the same three models and plotting conventions as
  Figure~\ref{fig:m82-floating-5kev}. The resonance energy is fitted
  in the right panel. The likelihood retains all forty bins, including
  bins with negative net counts.}
  \label{fig:m82-floating-1kev}
\end{figure}

We next fit the $1\,{\rm keV}$ spectrum extracted from the same
observations. This extraction differs from the released spectrum, so
the comparison changes more than the bin width.\footnote{Regrouping
the extracted $1\,{\rm keV}$ counts into $5\,{\rm keV}$ bins gives
$\chi^2=10.281,8.949,4.952$ for the power law, single-axion and
two-axion fits, respectively. The two-axion fit then has
$\omega_\star=35.03\,{\rm keV}$. These are a control for the extraction,
rather than the released-data results in Table~\ref{tab:m82-floating-5kev}.}
Figure~\ref{fig:m82-floating-1kev} and Table~\ref{tab:m82-floating-1kev}
show the three fits. The best two-axion point found has
$\omega_\star=39.3\,{\rm keV}$,
$m_H=3.70\times10^{-9}\,{\rm eV}$,
$m_L=1.74\times10^{-22}\,{\rm eV}$ and
$\gH=5.14\times10^{-11}\,{\rm GeV}^{-1}$.
The single-axion fit has $m_H=7.64\times10^{-10}\,{\rm eV}$ and
$\gH=1.06\times10^{-11}\,{\rm GeV}^{-1}$.
The two-axion fit lowers $\chi^2$ by $7.274$ relative to the power law and
$5.305$ relative to the single-axion fit. The preferred mass pairs therefore
differ between the two spectra.

\begin{table}[htbp]
  \centering
  \small
\begin{tabular}{@{}lrrrrr@{}}
\toprule
Model & $k$ & $\chi^2$ & $\nu$ & $\chi^2/\nu$ & $\Delta\chi^2_{\rm PL}$\\
\midrule
Power law only & 2 & 24.873 & 38 & 0.655 & 0.000\\
Single axion + power law & 4 & 22.905 & 36 & 0.636 & -1.968\\
Two axions + power law & 5 & 17.600 & 35 & 0.503 & -7.274\\
\bottomrule
\end{tabular}
  \caption{Goodness of fit for the forty $1\,{\rm keV}$ bins in
  Figure~\ref{fig:m82-floating-1kev}, with $\nu=40-k$.
  The parameter counts and $\Delta\chi^2$ definitions are the same as in
  Table~\ref{tab:m82-floating-5kev}.}
  \label{tab:m82-floating-1kev}
\end{table}
\FloatBarrier

We estimate how often fluctuations alone could give a fit improvement
at least as large as the one observed. In what follows we will follow a Frequentist's viewpoint by generating mock spectra. 
We do this separately assuming
that the power law, or the single-axion model, describes the true spectrum. We will try to quantify the chance of misidentifying a pure statistical fluctuation as the two-axion model resonance signal, \textit{i.e.} a false positive.
For either comparison model, denoted by $C$, the observed difference is
\begin{equation}
  \Delta\chi^2_{C,\rm obs}
  =\chi^2_{\rm two,\min}-\chi^2_{C,\min}.
  \label{eq:feature-observed-improvement}
\end{equation}
More negative values indicate a larger improvement from the two-axion model.
For example, the $5\,{\rm keV}$ fits give
$\Delta\chi^2_{\rm PL,obs}=-7.859$ and
$\Delta\chi^2_{\rm single,obs}=-6.311$.
The calculation proceeds as follows:
\begin{enumerate}
  \item Fit the comparison model to the data. Use its best-fit predicted
  counts as the mean of $N=10^4$ simulated spectra. In each bin, add an
  independent Gaussian fluctuation with standard deviation equal to the
  measured error $\sigma_i$; keep these errors fixed in all fits.
  \item Fit both the comparison model and the two-axion model to every
  simulated spectrum, and calculate their difference in minimum
  $\chi^2$, using Eq.~\eqref{eq:feature-observed-improvement}.
  Refit the power-law normalization and slope and, when present, the
  nonnegative axion amplitude. Search the saved signal spectra at different
  masses for each axion model, so the simulated fits have the freedom to
  select a feature at different energies.
  \item Count the simulations with at least the observed reduction in $\chi^2$,
  satisfying $\Delta\chi^2_{C,\rm sim}\leq\Delta\chi^2_{C,\rm obs}$.
  If this number is $N_{\rm tail}$, estimate the false-alarm probability by
  \begin{equation}
    p_C=\frac{N_{\rm tail}+1}{N+1}
        \simeq\frac{N_{\rm tail}}{N}.
    \label{eq:feature-simulation-pvalue}
  \end{equation}
  The added ones prevent a finite simulation from assigning zero
  probability when no simulated improvement exceeds the observed one.
\end{enumerate}
Thus $p_C$ is the probability, under the fitted comparison model and the
assumed errors, that statistical fluctuations yield a fit improvement at
least as large as observed after searching for a feature.
Repeating the search in every simulation accounts for the
freedom to choose its energy~\cite{Gross:2010LookElsewhere}.
We estimate this probability by simulation because a nonnegative signal
whose masses become irrelevant at zero strength does not satisfy the
conditions for the usual $\chi^2$ test based on the number of added
parameters~\cite{Protassov:2002SpectralFeatures}.

We express $p_C$ as a one-sided Gaussian significance, reported as a
number of $\sigma$. We choose this number so that the probability above
it in a standard normal distribution equals $p_C$: $\Phi^{-1}(1-p_C)$ in the unit of standard deviation,
where $\Phi$ is the standard normal cumulative distribution function.
For the $5\,{\rm keV}$ power-law comparison, $553$ of $10^4$ simulations
give $\Delta\chi^2\leq-7.859$. Hence
$p_{\rm PL}=554/10001=0.0554$, corresponding to $1.59\sigma$.
For the single-axion comparison, $293$ simulations give
$\Delta\chi^2\leq-6.311$, yielding $p_{\rm single}=294/10001=0.0294$
and $1.89\sigma$. For the $1\,{\rm keV}$ spectrum, the corresponding counts
are $2461$ and $2124$, at observed differences $-7.274$ and $-5.305$.
They give $p_{\rm PL}=0.246$ ($0.69\sigma$) and
$p_{\rm single}=0.212$ ($0.80\sigma$).

In the simulated fits, the continuous mass search is approximated by a
fixed library of saved signal spectra, and the power-law slope is searched
on a grid. The two-axion library samples
$29\leq\omega_\star\leq1000\,{\rm keV}$ and
$10^{-26}\leq m_L\leq5.6\times10^{-22}\,{\rm eV}$.
Resonance energies above the fitted band allow higher harmonics to
produce features within it. The finite library and the coarser numerical
integration used for some saved spectra make these approximate
probabilities for the stated search and fixed source, halo and detector
model. Holding the best-fit masses fixed instead gives local significances
of about $2.7\sigma$ relative to the power law in either spectrum.

While the fitted features remain tentative and do not establish an axion signal, we show that the resonance signal can potentially generate real features in X-ray observations, hence can be searched for in future X-ray missions. 

\FloatBarrier

\section{Discussion and conclusions}
\label{sec:conclusions}

In this work, we have shown that a light axion DM background can enhance the detectability of heavy axion species. This is achieved through the enhanced conversion probability between the propagating degrees of freedom, \textit{i.e.} the relativistic axion and photons traversing a static magnetic field. Unlike the usual AMR that requires a helical or harmonic magnetic profile, the DM-induced AMR can happen in a homogeneous magnetic background all thanks to the osillitory DM background. Despite the seemingly distinct resonance conditions, the induced parametric resonance can be understood using the exact same language as the AMR. On one hand, this again shows that AMR has a broad applicability, acrossing laboratory laser axion experiment \cite{Seong:2023AMR}, axion helioscopes \cite{Seong:2024Helioscopes}, and galaxies (this work); on the other hand, using the starburst galaxy M82, it provides an real-world example where AMR can also be relevant for DM signals.

Aside from the technical aspect of AMR, we show that, even without raising the axions' coupling strengths, certain part of the axiverse parameter space is bound to be easier to probe compared to the single-axion scenario. In this work, we show through the DM-induced AMR that there exists multiple strips in the $m_{aH} - m_{aL}$ plane for the pair of the two axion species where the star-emitted axions can generate more X-ray than the single-axion case with the same coupling strength, thanks to the resonance.

Lastly, let us briefly comment on the size of the angle $\theta_L$. Using Eq.~\eqref{eq:dm-background-field}, the peak value of the
light-axion field angle $\theta_L=a_L/f_{a_L}$ is
\[
  |\theta_L|_{\rm peak}
  =\frac{\sqrt{2\rho_{\rm DM}}}{m_L f_{a_L}}
  \simeq \frac{|\gL|\sqrt{2\rho_{\rm DM}}}{m_L},
\]
where the last expression assumes $|\gL| f_{a_L}\simeq1$.
For the local benchmark in Table~\ref{tab:local-m82-benchmark}, this
gives $0.053\,{\rm rad}$ ($3.0^\circ$). The corresponding photon
polarization rotation has half this amplitude, $0.026\,{\rm rad}$,
from Eq.~\eqref{eq:dm-rotation-angle}.
At the same central density, the best fits to the spectra with
$5\,{\rm keV}$ and $1\,{\rm keV}$ bins in
Section~\ref{sec:resonance-features} give field angles of $0.25$ and
$0.032\,{\rm rad}$, respectively. At fixed density and coupling, the
estimated angle grows as $1/m_L$ and reaches unity near
$m_L=6\times10^{-24}\,{\rm eV}$ for the central density and coupling used
in Section~\ref{sec:enhanced-constraints}.
Therefore, $a_L \ll f_L$ is maintained through the analysis.

\acknowledgments
The authors would like to thank Xiao-Jun Bi, Chao-Qiang Geng, Tao Liu, Joshua Ruderman, Yue-Lin Sming Tsai, and Yue-Liang Wu for useful discussions. CS is grateful for the invitation to the workshop "Axion-Photon Conversion: Terrestrial and Cosmic Searches" and the hospitality of Tsung-Dao Lee Center for Sciences and Arts (TDLCSA), where part of the work was done. 
CS and SW are partly supported by National Natural Science Foundation of China (NSFC) under Grant No. 12547104.
SY is supported by Basic Science Research Program through the National Research Foundation of Korea(NRF) funded by the Ministry of Education(No. RS-2026-25571203) and by IBS under the project code, IBS-R018-D1.

\appendix

\section{Dark-matter velocity dispersion in M82}
\label{app:velocity-dispersion}

The DM coherence scale in Eq.~\eqref{eq:m82-dm-domains} depends on the
spread of random velocities. We calculate this spread from the Jeans
equation, using the same mass model as the numerical notebook.
The gravitational field includes the entire DM halo, old stars, and gas.

\subsection{Mass model}
\label{app:m82-mass-model}

Our benchmark follows the pre-starburst initial mass budget of
Wang et al.~\cite{Wang:2024M82}. Its adopted parameters are
listed in Table~\ref{tab:m82-mass-model}.
This is a model choice for the gravitational potential; it is not a
new fit to the present-day mass distribution.
\begin{table}[htbp]
  \centering
  \begin{tabular}{@{}lll@{}}
    \toprule
    Component & Mass parameter & Scales\\
    \midrule
    DM halo & $M_{\rm vir}=6.0\times10^{10}M_\odot$
      & $R_{\rm vir}=53\,{\rm kpc}$, $r_c=0.5\,{\rm kpc}$\\
    Old stellar disk & $M_\star=3.3\times10^9M_\odot$
      & $a_\star=1.2\,{\rm kpc}$, $b_\star=0.2\,{\rm kpc}$\\
    Initial gas disk & $M_g=3.0\times10^9M_\odot$
      & $r_{1/2}=1\,{\rm kpc}$, $h=0.255646\,{\rm kpc}$\\
    \bottomrule
  \end{tabular}
  \caption{Mass model used in the Jeans calculation. The gas half-mass
  radius is defined inside a sphere and determines the scale $h$.
  The gas budget includes the precursors of recently formed nuclear stars.}
  \label{tab:m82-mass-model}
\end{table}

For the halo, we retain the smooth core-NFW enclosed mass in their
Eq.~(A8), with exponent $n=1$:
\begin{equation}
  M_h(<r)=M_{\rm vir}
  \frac{f(r/r_s)\tanh(r/r_c)}
       {f(R_{\rm vir}/r_s)\tanh(R_{\rm vir}/r_c)},
  \qquad f(x)=\ln(1+x)-\frac{x}{1+x}.
  \label{eq:appendix-halo-mass}
\end{equation}
The concentration relation in their Eq.~(A9),
$\log_{10}c_h=1.071-0.098[\log_{10}(M_{\rm vir}/M_\odot)-12]$,
gives $c_h=15.515$ and $r_s=R_{\rm vir}/c_h=3.416\,{\rm kpc}$.
This smooth profile differs from their subsequent constant-density
patch. The total DM halo density is obtained by differentiating
the enclosed mass,
\begin{equation}
  \rho_{\rm halo}(r)=\frac{1}{4\pi r^2}\frac{dM_h(<r)}{dr}.
  \label{eq:appendix-halo-density}
\end{equation}
For the adopted exponent $n=1$, the central limit is
$\rho_{\rm halo}(0)=3\rho_s r_s/(2r_c)$, where $\rho_s$ is the NFW
scale density.

Both disks use the Miyamoto--Nagai potential~\cite{Miyamoto:1975Disk},
\begin{equation}
  \Phi_{\rm MN}(R,z;M,a,b)
  =-\frac{GM}{\sqrt{R^2+(a+\sqrt{z^2+b^2})^2}},
  \label{eq:appendix-mn-potential}
\end{equation}
where $R$ is cylindrical radius. The old stellar disk is a single term
with the parameters in Table~\ref{tab:m82-mass-model}.
The total gas potential follows the three terms in Wang et al.'s
Table~A1:
\begin{equation}
  \begin{aligned}
    \Phi_g(R,z)&=\sum_{i=1}^3\Phi_{\rm MN}(R,z;w_iM_g,a_i,h),\\
    (a_1,a_2,a_3)/h&=(2.39,2.67,4.83),\\
    (w_1,w_2,w_3)&=(1.37,0.0779,-0.449)/0.9989.
  \end{aligned}
  \label{eq:appendix-gas-potential}
\end{equation}
The normalization corrects the rounding of the published coefficients.
The signed terms are summed before interpreting the gas density or mass.
We interpret the quoted gas half-mass radius as spherical;
$M_g(<1\,{\rm kpc})=M_g/2$ fixes $h$ in the table.
Nuclear precursors are included in $M_g$, without adding a separate
nuclear component.

\begin{figure}[htbp]
  \centering
  \includegraphics[width=0.76\textwidth]{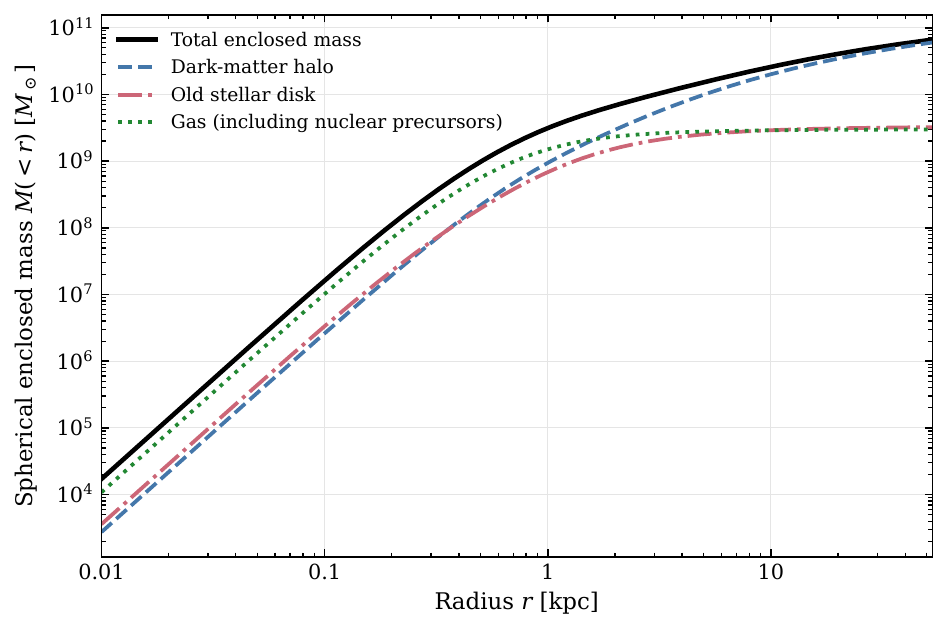}
  \caption{Enclosed mass of the adopted M82 model. Each curve gives the
  mass inside a sphere of radius $r$; the total includes all DM, old
  stars, and initial gas. The three signed gas basis terms are grouped
  into a single physical component.}
  \label{fig:m82-enclosed-mass}
\end{figure}

To use a spherical Jeans equation, we approximate the flattened
potential by its angular mean. With $\mu$ the cosine of the polar angle,
\begin{equation}
  \bar\Phi(r)=\frac12\int_{-1}^{1}d\mu\,
  \Phi_{\rm tot}\!\left(r\sqrt{1-\mu^2},r\mu\right),
  \qquad
  \frac{d\bar\Phi}{dr}=\frac{GM_{\rm tot}(<r)}{r^2}.
  \label{eq:appendix-monopole}
\end{equation}
The second equality follows from Gauss's law: the angular mean of the
radial gravitational field measures the physical mass inside a sphere,
including the flattened disks. Figure~\ref{fig:m82-enclosed-mass}
shows these masses.

\subsection{Jeans equation and numerical evaluation}
\label{app:m82-jeans}

We assume a stationary, nonrotating DM population with isotropic
random velocities. Then the dispersions of all three velocity
components coincide, $\sigma_r=\sigma_\theta=\sigma_\phi=\sigma_{1d}$,
and $\rho_{\rm halo}\sigma_{1d}^2$ acts as a pressure.
The first velocity moment of the collisionless Boltzmann equation
balances its gradient against gravity~\cite{Binney:2008Dynamics}:
\begin{equation}
  \frac{d(\rho_{\rm halo}\sigma_{1d}^2)}{dr}
  =-\rho_{\rm halo}\frac{d\bar\Phi}{dr}.
  \label{eq:appendix-jeans-equation}
\end{equation}
Integrating outward with vanishing pressure at infinity gives
\begin{equation}
  \sigma_{1d}^2(r)=\frac{G}{\rho_{\rm halo}(r)}
  \int_r^\infty ds\,
  \frac{\rho_{\rm halo}(s)M_{\rm tot}(<s)}{s^2}.
  \label{eq:appendix-jeans-solution}
\end{equation}
Thus the local dispersion depends on the gravitational potential and
tracer density at larger radii as well as at $r$.
We take $\rho_{\rm DM}=0.2\rho_{\rm halo}$ with the same spatial profile.
This constant fraction cancels between the numerator and denominator;
the full DM mass still contributes to $M_{\rm tot}$.

Numerically, we evaluate the angular mean with 128-point
Gauss--Legendre quadrature and tabulate the radial integrand on 8000
logarithmic nodes from $10^{-5}$ to $10^4\,{\rm kpc}$.
A reverse cumulative trapezoidal integral imposes zero pressure at
the outer node. We interpolate $\sigma_{1d}^2$ in $\ln r$ and then take
its square root; below the first node we use its resolved core value.
The halo is extrapolated beyond $R_{\rm vir}$, which sets its mass
normalization rather than a truncation radius.
Doubling the radial nodes, angular order, or outer radius changes
$\sigma_{1d}$ by less than $5\times10^{-6}$ fractionally along the
$0$--$25\,{\rm kpc}$ propagation ray.

\begin{figure}[htbp]
  \centering
  \includegraphics[width=0.76\textwidth]{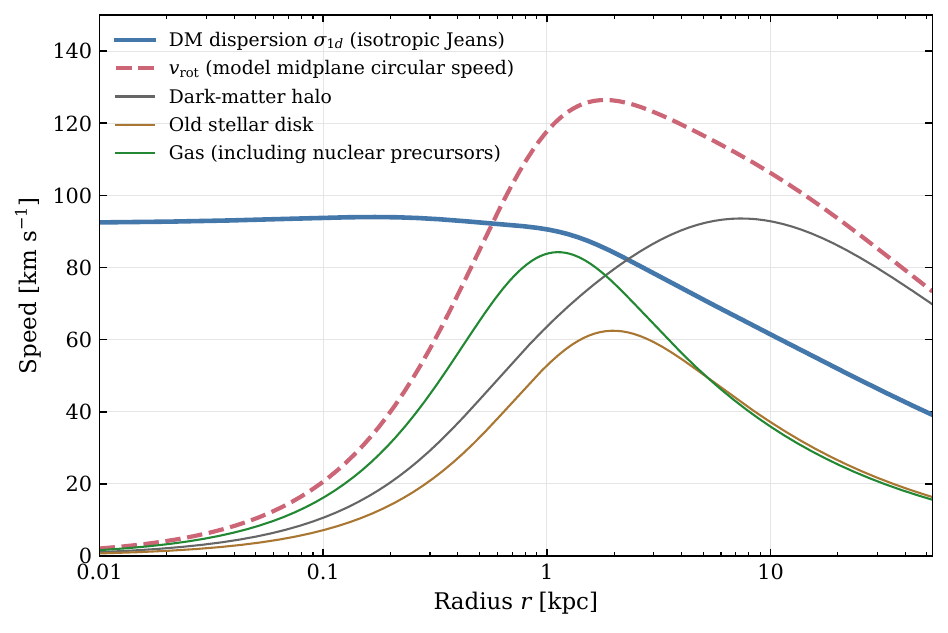}
  \caption{One-dimensional DM velocity dispersion from
  Eq.~\eqref{eq:appendix-jeans-solution}, overlaid with the midplane
  circular speed $v_{\rm rot}$ of the same mass model.
  Thin solid curves show the halo, old stellar disk, and total gas
  contributions, defined by $v_j^2=R\,\partial_R\Phi_j(R,0)$;
  their squares sum to $v_{\rm rot}^2$.
  Random motion provides pressure support and remains finite in the
  central core even where the circular speed vanishes.
  The dispersion determines the DM coherence length in
  Eq.~\eqref{eq:m82-dm-domains}.}
  \label{fig:m82-velocity-dispersion}
\end{figure}

For comparison, we define the model rotation speed by circular-orbit
balance in the disk midplane:
\begin{equation}
  v_{\rm rot}^2(r)
  =\left.R\frac{\partial\Phi_{\rm tot}(R,z)}{\partial R}
    \right|_{R=r,\,z=0}.
  \label{eq:appendix-rotation-speed}
\end{equation}
This uses the flattened potential, whereas the Jeans calculation uses
its spherical mean. Even in a spherical potential, pressure support
does not imply a universal ratio between random and circular speeds.
Figure~\ref{fig:m82-velocity-dispersion} compares the two: the finite
DM core gives $\sigma_{1d}(0)\simeq92.3\,{\rm km\,s^{-1}}$, while
$v_{\rm rot}$ tends to zero. Particles crossing the central region can
carry random velocities set by orbits extending to larger radii.
At $r=1$ and $25\,{\rm kpc}$, the dispersion is approximately
$90.6$ and $48.9\,{\rm km\,s^{-1}}$, respectively.

\subsection{Interference and DM wave-packet sizes}
\label{app:m82-wave-packets}

The Jeans equation fixes the smooth dispersion $\sigma_{1d}(r)$, but
does not specify a velocity-distribution shape or random scatter of
this dispersion. Here we assume Gaussian velocity components with
this fixed width and measure the resulting interference structure.
We then compute the overall amplitude of the DM. We measure the length for the amplitude to change by $\mathcal O(1)$, and use this length as a proxy for DM wave-packet sizes.

On the central ray, $r=z$. We use $\hbar=c=1$, with velocities expressed
in units of $c$, and write the real background field as
\begin{equation}
 a_L(z)=\frac{\sqrt{2\rho_{\rm DM}(z)}}{m_L}
       \operatorname{Re}\!\left[e^{im_Lz}A(z)\right].
 \label{eq:appendix-wave-envelope}
\end{equation}
The complex envelope $A(z)$ contains the sum of the velocity modes.
For a constant mode velocity $v_{z,j}$, removing the common mass
oscillation leaves the phase $-m_Lv_{z,j}z$ at leading Doppler order.
To follow the varying Jeans dispersion, we prescribe
\begin{equation}
 \begin{aligned}
 v_{z,j}(z)&=v_{z,j}(0)\frac{\sigma_{1d}(z)}{\sigma_{1d}(0)},\\
 A(z)&=\sum_j c_j\exp\!\left[-im_L\int_0^z v_{z,j}(z')\,dz'\right].
 \end{aligned}
 \label{eq:appendix-wave-modes}
\end{equation}
Thus a constant dispersion gives $A(z)=\sum_jc_j e^{-im_Lv_{z,j}z}$.
The spatial rescaling of the mode velocities is an additional
wave-model prescription; the Jeans calculation supplies the smooth
width. We neglect the kinetic-frequency correction of order $v^2$
for this size diagnostic.

At $z=0$, choose equally spaced velocity bins. Their shares of the
mean squared field amplitude follow the assumed Gaussian velocity
distribution:
\begin{equation}
 \begin{aligned}
 w_j&=\frac{\exp[-v_{z,j}^2(0)/(2\sigma_{1d}^2(0))]}
              {\sum_l\exp[-v_{z,l}^2(0)/(2\sigma_{1d}^2(0))]},\\
 c_j&=\sqrt{w_j/2}\,(X_j+iY_j).
 \end{aligned}
 \label{eq:appendix-wave-spectrum}
\end{equation}
All $X_j,Y_j$ are independent Gaussian numbers with zero mean and
unit variance, drawn once per realization and held fixed along the
ray. Hence $\langle|c_j|^2\rangle=w_j$ and
$\langle|A(z)|^2\rangle=\sum_jw_j=1$, where brackets average over
realizations. The square root converts the squared-amplitude weight
into a field amplitude. Interference of these modes produces the
fluctuating envelope without an additional scatter law for
$\sigma_{1d}$ or numerical smoothing of the real field.

Let us set up the amplitude-change criterion for the wave-packet proxy.
Starting at $z_0=0$, end an interval at the first point satisfying
\begin{equation}
 \bigl||A(z_{i+1})|-|A(z_i)|\bigr|=\eta A_{\rm rms},
 \qquad \eta=0.5,\qquad
 A_{\rm rms}=\sqrt{\langle|A|^2\rangle}=1.
 \label{eq:appendix-wave-amplitude-exit}
\end{equation}
Its physical length is $z_{i+1}-z_i$. The criterion acts on the
normalized envelope, so smooth changes in $\rho_{\rm DM}(z)$ do not
create boundaries. The fixed ensemble amplitude avoids division by
small local amplitudes. As checks, a common phase drift leaves these
boundaries unchanged, and a single plane wave never triggers one.
The next interval starts at the crossing amplitude. Every reference
therefore differs from $|A(0)|$ by an integer multiple of $\eta$;
this starting-value convention remains part of the definition.

We include some more details for the numerical sampling we performed. We first construct 512 envelopes in a uniform
reference region with dispersion $\sigma_{1d}(0)$. The velocity grid
has $|v_{z,j}(0)|\leq8\sigma_{1d}(0)$ and spacing
$2\pi\sigma_{1d}(0)/1024$, corresponding to a spatial period
$1024\ell_{\rm DM}(0)$. We sample to $512\ell_{\rm DM}(0)$ with
step $\ell_{\rm DM}(0)/128$, and retain complete intervals starting
between $64\ell_{\rm DM}(0)$ and $192\ell_{\rm DM}(0)$.
Linear interpolation of $|A|$ locates the crossings. This gives
36,690 reference intervals at $\eta=0.5$, with mean length
$1.788\ell_{\rm DM}(0)$ and 16th--84th percentile range
$(0.409$--$3.114)\ell_{\rm DM}(0)$.
Comparing $\eta=0.25$ and $1$ checks the threshold dependence.
Halving the spatial step for 64 matched fields changes the length
percentiles by less than $6\times10^{-5}$ fractionally at all three
thresholds.
\begin{figure}[htbp]
  \centering
  \includegraphics[width=0.66\textwidth]{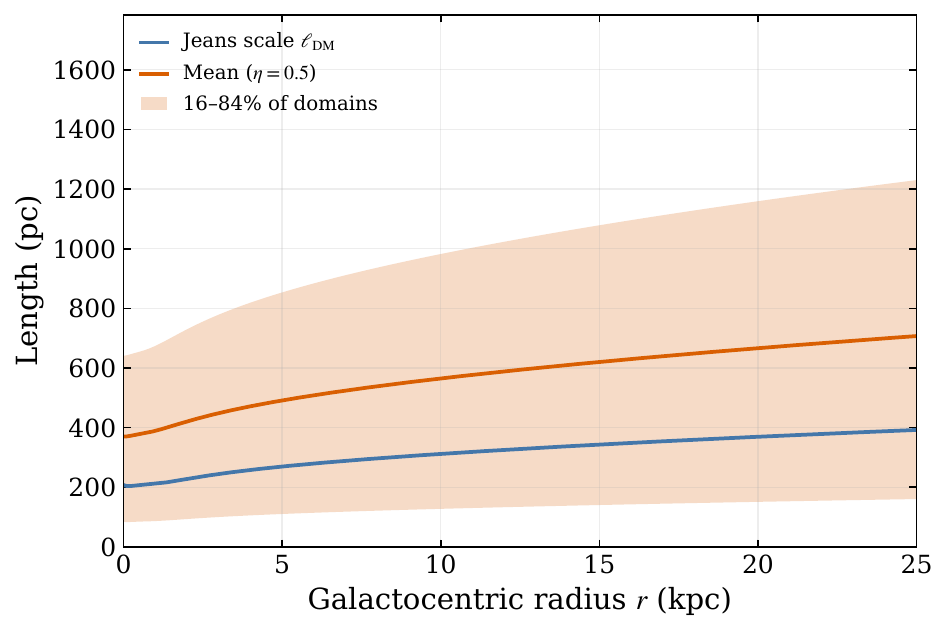}
  \caption{Amplitude-change domains for $m_L=10^{-22}\,{\rm eV}$, measured in 512 Gaussian wave-envelope realizations, with the M82 Jeans dispersion fixed. A domain ends when $|A|$ first differs from its starting value by $0.5A_{\rm rms}$. The blue curve is the Jeans reference $\ell_{\rm DM}=1/(m_L\sigma_{1d})$. Orange shows the mean and 16th--84th percentile range for domains starting at $r=z$ on the central ray. The mean is computed after mapping each measured length to physical distance. These amplitude features do not imply independent phase resets.}
  \label{fig:m82-dm-domain-size-wave}
\end{figure}

Finally, we take into account that $\sigma_{1d}$ has a $z$ dependence. The same velocity rescaling maps a measured reference interval
$L_{\rm ref}$ to a physical length $L(z)$ starting at $z$ through
\begin{equation}
 \int_z^{z+L(z)}\sigma_{1d}(z')\,dz'
   =\sigma_{1d}(0)L_{\rm ref}.
 \label{eq:appendix-wave-length-mapping}
\end{equation}
Figure~\ref{fig:m82-dm-domain-size-wave} shows the mean and
16th--84th percentile range of these physical lengths, with the
Jeans dispersion held fixed. In this modeling, the complex field remains continuous across the
amplitude boundaries, with no independent phase resets unlike the simplified DM domain model adopted in the main text. Therefore, the mean derived from this procedure does not need to coincide with
$\ell_{\rm DM}$, yet they are quite consistent, justifying our adoption of the DM wave packets in the main text. 

\FloatBarrier

\bibliographystyle{JHEP}
\bibliography{references}

\end{document}